\documentclass{svjour3}
\smartqed
\usepackage{graphicx}
\usepackage{amsfonts}
\usepackage{amssymb}

\begin{document}

\title{Testing  the Concept of  Quark-Hadron Duality  with the
ALEPH $\tau$ Decay Data}

\titlerunning{Testing  the Concept of  Quark-Hadron Duality}        

\author{B.A. Magradze}

\institute{ B.A. Magradze \at
            Andrea  Razmadze Mathematical Institute of I. Javakhishvili Tbilisi State University,
             2, University st., 0186 Tbilisi, Georgia  \\
                            \email{magr@rmi.acnet.ge}}


\date{Received: date / Accepted: date}
\maketitle

\begin{abstract}
We propose a modified procedure for extracting the numerical value
for the strong coupling constant $\alpha_s$ from the $\tau$ lepton
hadronic decay rate into non-strange particles in the vector
channel. We employ the concept of the quark-hadron duality
specifically, introducing a boundary energy squared $s_{\rm p}>0$,
the onset of the perturbative QCD continuum in Minkowski space
\cite{BLR,Rafa,PPR}. To approximate the hadronic spectral function
in the region $s>s_{\rm p}$, we use Analytic Perturbation Theory
(APT) up to the fifth order. A new feature of our procedure is
that it enables us to extract from the data simultaneously  the
QCD scale parameter $\Lambda_{\overline{\rm MS}}$ and the boundary
energy squared $s_{\rm p}$.  We carefully determine the
experimental errors on these parameters which come from the errors
on the invariant mass squared distribution. For the $\overline{\rm
MS}$ scheme coupling constant,   we obtain
$\alpha_s(m^{2}_{\tau})=0.308\pm 0.014_{\rm exp.}$. We show that
our numerical analysis is  more stable against higher-order
corrections than the standard one. The extracted value for the
duality point $s_{\rm p}$ is found  surprisingly stable against
perturbation theory corrections $s_{\rm d}= 1.71\pm 0.05_{\rm
exp}\pm 0.00_{\rm th}\,\, {\rm GeV^{2}}$.Additionally, we
recalculate the ``experimental" Adler function in the infrared
region using final ALEPH results. The uncertainty on this function
is also determined.

\end{abstract}

\section{Introduction}
\label{intro}
 The hadronic $\tau$ decays serves as an ideal
laboratory for testing quantum chromodynamics  (QCD) in a
relatively low energy regime. In the past, various techniques
(fixed order perturbation theory, contour improved perturbation
theory, effective charge approach, renormalons, dispersive
approach) have been devised to improve the reliability of the
predictions of the theory for the $\tau$ system. In this boundary
area of the energy, perturbative ideas are still applicable due to
relatively large mass of the $\tau$ lepton, while non-perturbative
effects are expected to be small. Usually, they are under control
within Wilson Operator Product Expansion (OPE) \cite{shifman}. It
is known that the main calculational tool in perturbative QCD
(pQCD) the renormalization group improved perturbation theory
augmented with the OPE can not be used locally in the time-like
region even at high energy. Fortunately, this problem has been
resolved in earlier work \cite{PQW} by means of the idea of the
quark-hadron duality. This enabled one to employ the QCD
perturbation theory in Minkowski region  to calculate some global
(inclusive) quantities like $\tau$ lepton decay rate. Although the
quark-hadron duality cannot be justified  rigorously from the
first principles, in practice this idea works good enough. Using
the duality,   an accurate description of the $\tau$ lepton decay
data was achieved (see the seminal work \cite{BNP} and the
literature therein). However, one should always keep in mind that
the duality between a physical quantity and its quark-gluon
perturbation theory representation is only approximative and thus
it must inevitable be violated (see the review \cite{shifman1} and
the literature therein). To identify general mechanism of possible
Duality Violations (DVs), special QCD inspired models for the
hadronic spectral functions (e.g. the instanton-based and
resonance-based models \cite{shifman1}  as well as the models
motivated by the large $N_{c}$ limit of the theory \cite{PPhR})
have been studied. In these models DVs in fact occur. Presumably,
DVs arise due to the lack of the convergence of the OPE on the
Minkowski axis. If this is the case, then  the analytical
continuation of the truncated OPE series from the Euclidean region
to the physical axis  is questionable \cite{shifman1}.

 In recent years, the accuracy of the measurements of the
observables of the $\tau$ lepton system has been essentially
improved (for the recent results of the ALEPH collaboration see
\cite{ALEPH,compilation,DHZ1,DHZ}). This enables one to extract
the parameters of the standard model from $\tau$ data with very
high precision. Of particular interest is the numerical value of
the strong coupling constant $\alpha_s$. Admittedly, one of the
most precise determinations of the strong coupling constant comes
from the analysis of the $\tau$ data (for most recent results see
\cite{DHZ}). An independent low-energy highest-precession
determination of $\alpha_s$ comes from lattice QCD simulations
combined with experimental data for hadron masses \cite{Mason}.
These two highest-precision determinations extrapolated to the Z
mass yield
\begin{eqnarray}
\alpha_{s}(M_{z}^{2})&=& 0.1212\pm 0.0011\quad(\tau\quad \rm{decay})\\
\label{latt} \alpha_{s}(M_{z}^{2})&=& 0.1170\pm
0.0012\quad(\rm{lattice}).
\end{eqnarray}
Note that the agreement between these two results, with the errors
quoted, is not good. They differ from each other by about 2.6
standard deviations. Furthermore, the lattice determination is
closer to $\alpha_{s}(M_{z}^{2})$ values obtained from high energy
experiments. Thus, the reliability of the estimates from the
$\tau$-lepton data has been called in question
\cite{DHZ,Malt,Domin,Domin1,Pich,Nari,CGP1,CGP,Boito}. The small
but still significant non-perturbative effects have been included
into analysis \cite{Nari,Boito}. On the one hand, the impact of
the  higher order terms of the OPE (neglected in the standard
analyzes)   has been estimated \cite{Malt,Domin,Domin1,Pich}. It
was confirmed that their influence on the extracted value of
$\alpha_s$ is not small in the separate vector and axial vector
channels. To suppress these contributions in the finite energy sum
rule   the so-called pinched weights introduced
\cite{Malt,Domin,Domin1,Pich}. An independent estimation of
possible non-perturbaive corrections to the finite energy sum rele
(direct instantons, duality violation and tachyonic gluon mass)
which cannot be described within OPE can be found in \cite{Nari}.
To estimate systematic effects from DVs, recently  the authors of
\cite{DHZ} have analyzed the ALEPH $\tau$ data for the V+A
spectral function using two different models of DVs. These models
were previously considered in \cite{shifman1}.    It was confirmed
(within this models) that DVs effects in this channel  are
completely negligible. However, this problem has been reconsidered
in \cite{CGP}. There the separate vector (V) and axial-vector (A)
spectral data have  been analyzed. To describe DVs coming from the
region $s\geq 1.1 \,\,{\rm GeV^{2}}$ physically motivated models
for these spectral functions have been suggested. Analyzing the
$\tau$ data provided by the ALEPH collaboration, the authors of
\cite{CGP} have concluded that DVs are not small. An additional
systematic error in the value of the coupling constant coming from
DVs  has been estimated on the level
$\delta\alpha_{s}(m_{\tau}^{2})\approx 0.003-0.010$.

 As is well known,  in the time-like region the renormalization group
(RG) invariance cannot be used unambiguously. Usually, the QCD
corrections to the $\tau$ lepton decay rate $R_{\tau}$ is
expressed  via the contour integral of the associated Adler
function multiplied by the known weight function. This
representation is valid owing to special analyticity structure of
the corresponding  exact current-current correlation function. The
Adler function is represented via the truncated perturbation
theory series and the integral is taken over the circle of radius
$m_{\tau}^{2}$ ($m_{\tau}$ stands for the $\tau$-lepton mass) in
the complex energy squared plane \cite{BNP}. One possibility is to
integrate term-by-term the truncated perturbation theory series
over the contour and then perform the RG improvement. This
approach is referred to as fixed order perturbation theory (FOPT).
Alternatively, one can insert the RG improved truncated series for
the Adler function inside the contour integral and then perform
the integral. This approach  suggested in \cite{pivo-0,pivo,DP}
was termed contour improved perturbation theory (CIPT). The
advantage of  CIPT is that it enables to resume some higher-order
contributions to the rate. These two approaches lead to differing
results. The values of $\alpha_s$ extracted from $\tau$ decays
employing CIPT have always been higher. A detailed comparison of
these two approaches may be found in recent works \cite{J,BJ}. A
practical review of various approaches to the $\tau$ decay rate
may be found in \cite{DHZ1}.

The inclusive quantity like hadronic $\tau$ decay rate may be
accurately  described within pure perturbative approach, provided
the DVs are small. Indeed, in the V+A channel, the nonperturbative
power suppressed contributions described by the OPE (continued
analytically to the time-like region) have been estimated to be
small \cite{BNP,DHZ,Nari}. However, the large value of the running
coupling parameter at the $\tau$ lepton mass scale leads to the
large renormalization scheme dependence of perturbative
predictions.  To reduce this dependence various resummation
techniques have been developed (see, for example,
\cite{KKP,KAST,raczka}). In \cite{KKP}, the V+A $\tau$-lepton
decay data was analyzed within a modified extraction procedure
based on the effective charge approach. The numerical analysis has
been performed in the internal renormalization scheme of the
$\tau$ system and then the result was translated into the
$\overline{\rm MS}$ scheme using renormalization scheme
transformation. This procedure yields smaller value for the
coupling constant. Similarly, in \cite{KAST} and \cite{raczka} in
calculations of the $\tau$ decay rate the minimal sensitivity and
effective charge schemes were  used. In this  way the reliability
of the estimates for the coupling constant has been improved.

A serious shortcoming of the conventional  perturbation theory
approximations to the current-current correlation functions
parameterized in terms of the running coupling is that they do not
obey  correct analytical properties of the corresponding exact
quantities. The analytical properties are violated due to  the
non-physical Landau singularities of the perturbative running
coupling which appear at small space-like momenta (for the
analytical structure of the perturbative coupling beyond the
one-loop order see \cite{ggk,my1,my2,my3}). Supposedly, these
singularities may deteriorate the extracted values of the
parameters \cite{pivo1}. This problem does not arise within
dispersive or analytic approaches to  pQCD. At present, several
such approaches are being intensively developed
\cite{SS,DMW,grunb,MSS,Y,SS1,S1,Shirk,SS2,MSa,MSS1,S2,cvetic,cvetic1,baku,baku1,baku2,prosp}.
In works \cite{MSS} and \cite{Y}, the $\tau$ lepton decay rate has
been analyzed within a simple and effective dispersive technique,
the Analytic Perturbation Theory (APT) (for reviews see
\cite{SS1,S1,SS2,S2,prosp}). However, the minimal analytic QCD
model (the same APT) predicts, from the non-strange $\tau$ lepton
decay data, too large value for the strong coupling constant,
$\alpha_s(m_{\tau}^2)=0.403\pm 0.015$ \cite{Y}. The advantages and
shortcomings of the three approaches to the $\tau$ decays (FOPT,
CIPT and APT) were thoroughly analyzed in \cite{MSa}. It should be
noted that the APT as well as its generalized versions suggested
more later \cite{cvetic,cvetic1,baku,baku1} proved to be very
useful from the phenomenological point of view.
  A remarkable feature  of these modified expansions is the better
convergence and improved stability property with respect to change
of the renormalization scheme. Nevertheless, one should keep in
mind that an analytic approach based only on perturbation theory
can not be defined unambiguously, since there is not a unique
recipe for removing the Landau singularities from the running
coupling.

A particular problem emerges from the observation that the QCD
perturbation theory augmented with the OPE fail  to describe the
detailed infrared behavior of the  Adler function associated with
the $\tau$ decay rate \cite{PPR}. To treat this problem a more
general framework is required. A suitable theoretical framework
was suggested in \cite{PPR}. There the hadronic non-strange vector
spectral function $v_{1}(s)$ \footnote{ We use the normalization
of the spectral function with the naive parton prediction
$v_{1,\rm naive}=1/2$. }  was represented by a simple {\it ansatz}
\begin{equation}
\label{ansatz1 } v_{1}(s)\approx\theta(s_{\rm p}-s) v_{1}^{\rm
np.}(s)+\theta(s-s_{\rm p})v_{1}^{\rm pQCD}(s),
\end{equation}
where $v_{1}^{\rm pQCD}(s)$ is the perturbation theory
approximation to the spectral function and  $s_{\rm p}$ is the
onset of perturbative continuum \footnote{The inequality $0<s_{\rm
p}<m_{\tau}^{2}$ is assumed.}, an infrared boundary in Minkowski
region above which we trust pQCD. The non-perturbative component
of the spectral function $v_{1}^{\rm np.}(s)$  was described by a
resonance based model (``the lowest meson dominance approximation
to large-$N_{c}$ QCD"). Using this model the authors of \cite{PPR}
have achieved correct matching in the intermediate region between
the pQCD and Chiral Perturbation Theory predictions for the Adler
function \footnote{ The infrared behaviour of the Adler function
was also correctly described within APT \cite{MSS1}. However, to
reproduce the $\tau$ data, APT requires large effective quark
masses.}. To compare the Adler function evaluated from
(\ref{ansatz1 }) to the experiment the authors of \cite{PPR} have
also constructed the ``experimental" spectral function
\begin{equation}
\label{GD} v_{1}^{\rm ``exp"}(s)= \theta(s_{\rm p}-s) v_{1}^{
 \rm exp}(s)+\theta(s-s_{\rm p})v_{1}^{\rm pQCD}(s),
\end{equation}
where $ v_{1}^{\rm exp}(s)$ is the genuine experimental part of
the total ``experimental'' spectral function which  is measured
with high precision by ALEPH \cite{ALEPH,ALEPH1} and OPAL
\cite{OPAL} collaborations in the range
$0<\sqrt{s}<m_{\tau}=1.777\, GeV$. Formula (\ref{GD}) extends the
spectral function beyond the range accessible in  the experiment.
Formulas (\ref{ansatz1 }) and (\ref{GD}) provide   practical
realizations of the concept of the quark-hadron duality (see the
original works \cite{BLR,Rafa}). The {\it ansatz}  (\ref{ansatz1
}) may be considered as an alternative for the  truncated OPE in
Minkowski region. The conventional formulation of the duality may
be recovered from formulas (\ref{ansatz1 }) or (\ref{GD}) by
taking the limit $s_{\rm p}\rightarrow 0$ and introducing the OPE
contributions \footnote{Strictly speaking this is true if the
perturbation theory component of the spectral function $v_{1}^{\rm
pQCD}(s)$ is evaluated within FOPT or
 APT.}. Note that,   the non-perturbative corrections to the
spectral function described by model (\ref{ansatz1 })  are
essentially confined in the low energy region $0<s<s_{\rm p}$.

In this paper we concentrate on  formula (\ref{GD}). Our aim is to
utilize the total information encoded in this representation. We
recall that the authors of  \cite{PPR} have used {\it ansatz}
(\ref{GD}) to extract  the numerical value for the parameter
$s_{\rm p}$ from the experimental data.   For the $\overline{\rm
MS}$ scheme scale parameter (for the three active flavours)  they
used the estimate
\begin{equation}
\label{L-PPR} \Lambda_{\overline{\rm MS}}=(372\pm 72)\,{\rm MeV}.
\end{equation}
The QCD component   of the spectral function, $v_{1}^{\rm
pQCD}(s)$, was determined   from the order ${\cal
O}(\alpha_{s}^{3})$ approximation to the Adler function.   The
approximation was constructed in terms of the exact numeric
two-loop running coupling constant, normalized at the scale
$s_{\rm p}$. The experimental component $v_{1}^{\rm exp.}(s)$ was
reconstructed from the ALEPH collaboration data obtained in 1999
\cite{ALEPH1}. Note that the estimate (\ref{L-PPR}) is close to
the ALEPH result for the scale parameter obtained  for that time
 $$\Lambda_{\overline{\rm MS}}=(370\pm
 51)\,{\rm MeV}.$$
However, these two results for $\Lambda_{\overline{\rm MS}}$
should not be compared. The final result of the collaboration for
the coupling constant corresponds to the average of the two values
obtained within the FOPT and CIPT approaches, while authors of
\cite{PPR} used only FOPT. Furthermore, in the ALEPH analysis the
estimate for ${\cal O}(\alpha_{s}^{4})$ term was also included,
while the QCD scale parameter was extracted using the exact
(numeric) four-loop running coupling. Using  the {\it ansatz}
(\ref{GD}) the authors of \cite{PPR} have derived consistency
condition from the OPE , an equation relating the parameters
$s_{\rm p}$ and $\Lambda_{\overline{\rm MS}}$. From this equation,
with the estimate (\ref{L-PPR}), they have found that
\begin{equation}
\label{s-b-PPR} s_{\rm p}=(1.60\pm 0.17)\,{\rm GeV}^{2}.
\end{equation}
Usually, it is  more convenient to compare the time-like
experimental data with theory via the Adler function, the object
determined in the space-like region \cite{EJKV}\footnote{we use
notation $q^{2}=-Q^{2}$ and $Q^{2}>0$ for space-like momenta}
\begin{equation}
\label{Adler}
D(Q^{2})=Q^{2}\int_{0}^{\infty}{2v_{1}(s)ds\over(s+Q^{2})^{2}},
\end{equation}
for this quantity reliable approximations are constructed in pQCD,
in massless \cite{ChKT,GKL,Surgul,Baikov,chetyrk2} as well as in
massive cases \cite{chetyrk2,EJKV}. The ``experimental'' Adler
function is obtained by inserting {\it ansatz} (\ref{GD}) into
integral (\ref{Adler})
\begin{equation}
\label{Exp}  D_{\rm ``exp"}(Q^{2})= { D}_{\rm exp}(Q^{2},s_{\rm
p})+ D_{\rm pQCD}(Q^{2},s_{\rm p}),
\end{equation}
where the  experimental and perturbation theory components of the
total ``experimental" Adler function are defined as
\begin{equation}
\label{Adlerp} D_{\rm exp}(Q^{2},s_{\rm p})=Q^{2}\int_{0}^{s_{\rm
p}}{2 v_{1}^{\rm exp}(s)d\,s\over(s+Q^{2})^{2}},\qquad D_{\rm
pQCD}(Q^{2},s_{\rm p})=Q^{2}\int_{s_{\rm p}}^{\infty}{2v_{1}^{\rm
pQCD}(s)d\,s\over(s+Q^{2})^{2}}.
\end{equation}
Note that the ``experimental'' Adler function   is not wholly
experimental quantity, since it depends also on the theoretical
component $ D_{\rm pQCD}(Q^{2},s_{\rm p})$. The latter may be
calculated using different theoretical approaches. For example,
one may apply FOPT or APT. Furthermore, the result will depend on
the higher order corrections to the $\beta$-function and to the
Adler function. In the past years, the ``experimental'' Adler
function was employed for testing various theoretical
approximations to the Adler function \cite{PPR,MSS1,cvetic}.

In view of appearance of final ALEPH  data in 2005
\cite{ALEPH,compilation} it is worthwhile   to recalculate the
``experimental'' Adler function. In this paper,  we will use
different strategy for extracting numerical values of the
parameters from the data. The distinguishing feature of our
analysis is that we will determine both parameters
($\Lambda_{\overline{\rm MS}}$ and $s_{\rm p}$) self-consistently.
Secondly, we pay particular attention to the estimation of the
experimental errors on the parameters and Adler function.
Furthermore, we will use a dispersive approach \footnote{The
difference between our framework and   APT  of Shirkov and
Solovtsov is clarified in Sect.~2.}.

In Sect. 2 we evaluate the perturbative component of the hadronic
spectral function  up to  order ${\cal O}(\alpha_{s}^{5})$ within
the dispersive approach. Then, we derive a transcendental system
of equations for the parameters $\Lambda_{\overline{\rm MS}}$ and
$s_{\rm p}$. The first equation of the system follows from the OPE
for the current-current correlation function in the limit of
massless quarks.  The second equation for the parameters is a
consequence of  the quark-hadron duality implemented by means of
the {\it ansatz} (\ref{GD});   perturbation theory is used to
calculate the decay rate of the $\tau$-lepton  into hadrons of
invariant mass larger than $\sqrt{s_{\rm p}}$. In Sect. 3 we solve
the system of equations for the parameters numerically. To
determine the empirical contributions in these equations, we
employ the final ALEPH data on the non-strange vector invariant
mass squared distributions which are available in
\cite{compilation}. To test the stability of the numerical results
against the QCD perturbative corrections, we use different
approximations to the Adler function from order ${\cal
O}(\alpha_{s})$ to order ${\cal O}(\alpha_{s}^{5})$. This enables
us to determine the  indicative theoretical errors \cite{KKP} on
the extracted numerical values of the parameters.  Our approach,
which we refer to as ${\rm APT}^{+}$, is compared with the
standard CIPT. In the most of the calculations, we use the
four-loop running coupling. In Sect. 4, we present numerical
results for the ``experimental'' Adler function obtained from the
final ALEPH data. The  values and associated experimental errors
of the function are tabulated in the region $Q=0-1.5\,\,{\rm
GeV}$. Our conclusions are given in Sect. 5. In Appendix A we give
some practical formulas obtained from the explicit (series)
solution  to the higher order RG equation. The statistical errors
on the parameters are carefully estimated in Appendix B. In
Appendix C we present some required results obtained within
standard CIPT.

\section{Theoretical Framework}
The main quantity of interest for following analysis is the Adler
function associated with the vector current two-point correlator.
The perturbative expansion of this function in the limit of
vanishing quark masses reads \cite{J}
\begin{equation}
\label{expansion1} D(Q^{2})=
\sum_{n=0}^{\infty}a^{n}_{s}(\mu^{2})\sum_{k=1}^{n+1}kc_{n,k}L^{k-1}\quad{\rm
where} \quad L\equiv \ln{Q^{2}\over \mu^{2}},
\end{equation}
$a_{s}(\mu^{2})={\alpha_{s}(\mu^{2})\over \pi}$ with
$\alpha_{s}(\mu^{2})$ being the strong coupling constant
renormalized at the scale $\mu$. Since the Adler function is a
physical quantity, it satisfies a homogenous RG equation. This
fact enables us to choose $\mu^{2}=Q^{2}$. Then the expansion
(\ref{expansion1}) may be reexpressed as an asymptotic expansion
in powers of the running coupling $\alpha_{s}(Q^{2})$
\begin{equation}
\label{expansion2} D_{\rm
RGI}(Q^{2})=\sum_{k=0}^{\infty}d_{k}\left({\alpha_{s}(Q^{2})\over\pi}\right)^{k},
\end{equation}
where $d_{n}=c_{n,1}$ and the subscript ``RGI'' refers to the
renormalization group improved perturbation theory. The first two
coefficients in series (\ref{expansion2}) are universal
$d_{0}=d_{1}=1$. The coefficients of order $a_{s}^{2}$ and
$a_{s}^{3}$ in the $\overline{\rm MS}$ scheme have been calculated
about thirty and fifteen years ago \cite{ChKT,GKL,Surgul}.
Recently, the authors of \cite{Baikov} have calculated the
coefficient $d_{4}$ in the case of massless quarks by using
powerful computational techniques. The known higher order
coefficients  in the $\overline{\rm MS}$ scheme for $n_f=3$ quark
flavours take values $d_{2}\simeq 1.6398$, $d_{3}\simeq 6.3710$
and $d_{4}\simeq 49.0757$.

 In practice the series (\ref{expansion2}) should be truncated. The
obtained approximations to the  Adler function do not obey correct
cut-plane analyticity properties of the exact function because of
the non-physical ``Landau singularities'' which present in the
perturbative running coupling.  The exact Adler function $D(z)$
($z=Q^{2}=-q^{2}$) is known to be analytic except the cut running
along the negative real axis. This fact enables us to calculate
the hadronic non-strange vector spectral function from the  Adler
function via the contour integral
\begin{equation}
\label{spectral}
 v_{1}(s)={1\over 4\pi\imath}\oint_{-s-\imath
0}^{-s+\imath 0}{D(z)\over z}d\,z,
\end{equation}
where the path of integration, connecting the points $-s\mp\imath
0$ on the complex $z$-plane,   avoids the  cut running along the
real negative axis. The integral being traversed in a positive
(anticlockwise) sense. In this paper we shall assume, without loss
of generality, that the approximation (\ref{expansion2}) to the
Adler function has only one non-physical singularity located on
the positive real axis. This is the case, for  the exact
(explicitly solved) two-loop order running coupling in
$\overline{\rm MS}$ like renormalization schemes \footnote{The
analytic structure of the explicit exact solution to the RG
equation at the two-loop order has been determined in
\cite{ggk,my1,my2}.}. On the other hand,  a running coupling at
higher orders may be expanded in powers of the exact (explicitly
solved) two-loop order coupling \cite{kour,join}
\begin{equation}
\label{sersol} \alpha_{s}^{(\rm
k-loops)}(Q^{2})=\sum_{n=1}^{\infty}{\cal
C}_{n}^{(k)}\alpha_{s}^{(\rm two-loops)n}(Q^{2})|_{\rm exact},
\end{equation}
where the numerical coefficients ${\cal C}_{n}^{(k)}$ are
determined in terms of the $\beta$-function coefficients (see
Appendix A). It was shown in \cite{my3} that this series has a
sufficiently large radius of convergence in the space of the
coupling constants,  and its partial sums  provide very accurate
approximations to the exact k-th  order ($k>2$) coupling in the
complex $Q^{2}$ plane. To construct accurate approximations to the
running coupling for small values of $|Q|^{2}$, one should keep
sufficiently large number of terms in the partial sum. The Adler
function evaluated with this approximation to the coupling has
only one non-physical singularity located on the positive
$Q^{2}$-axis. The corresponding cut runs along the finite interval
of the positive $Q^{2}$-axis. Nevertheless, formula
(\ref{spectral}) is still valid provided that the integration
contour avoids   the physical as well as  non-physical cut.

Let us separate  out the parton level term from the perturbative
Adler function
\begin{equation}
\label{RGI_PT} D_{\rm RGI}(Q^{2})=1+d_{\rm RGI}(Q^{2}): \quad
d_{\rm RGI}(Q^{2})=\sum_{k=1}^{\infty}d_{k}a_{s}^{k}(Q^{2}),
\end{equation}
where $a_{s}(Q^{2})={\alpha_{s}(Q^{2})/\pi}$. As it was discussed
above, the function $d_{\rm RGI}(Q^{2})$ is analytic except the
cuts running along the real $Q^{2}$-axis. The  physical cut runs
along the real negative  semi-axis $ -\infty<Q^{2}<0$, and the
non-physical cut runs along the positive interval $ 0<Q^{2}<s_{\rm
L}$, where the point $Q^{2}=Q^{2}_{\rm L}\equiv s_{\rm L}>0$
corresponds to the ``Landau singularity''. We may then write a
Cauchy relation
\begin{equation}
\label{cauchy} d_{\rm RGI}(Q^{2})={1\over
2\pi\imath}\oint_{\Gamma} {d_{\rm RGI}(w)\over {w-Q^{2}}}d\,w
\end{equation}
where the integral is taken round the closed contour $\Gamma$
drawn in Fig.1.  The contour  consists of the arc of the circle
$|Q^{2}-s_{\rm L}|=s_{\rm L}$, straight lines parallel to the real
negative $Q^{2}$ axis and passes round a big circle. Using formula
(\ref{cauchy}) together  with the asymptotic condition $d_{\rm
RGI}(z)\rightarrow 0$ as $|z|\rightarrow\infty$, we derive a
violated dispersion relation (DR)
 \begin{figure}
\centering
\includegraphics[scale=1]{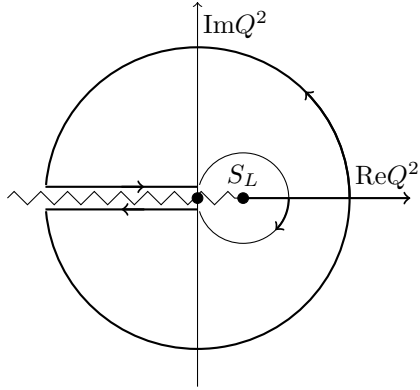}
\caption{Contour in the complex $Q^{2}$ plane used in the Cauchy relation (\ref{cauchy}).
 Branch points on the real axis are represented by the blobs and brunch cuts
 by the zigzagging  line.}

\end{figure}
\begin{equation}
\label{VDR} d_{\rm RGI}(Q^{2})=d_{\rm APT}(Q^{2})+d_{\rm L}(Q^{2})
\end{equation}
here the function $d_{\rm APT}(Q^{2})$ satisfies the  DR
\begin{equation}
\label{DR} d_{\rm
APT}(Q^{2})={1\over\pi}\int_{0}^{\infty}{\rho_{\rm
eff}(\sigma)\over{\sigma+Q^{2}}}d\,\sigma,
\end{equation}
with the effective spectral density
\begin{equation}
\label{specfun} \rho_{\rm eff}(\sigma)={\rm Im}\{d_{\rm
RGI}(-\sigma-\imath 0)\}.
\end{equation}
It is to be noted here that the function
\begin{equation}
\label{Aimage}
 D_{\rm APT}(Q^{2})=1+d_{\rm APT}(Q^{2})
\end{equation}
is the analytic image of the perturbative Adler function
determined in the sense of the Analytic Perturbation Theory (APT)
approach of Shirkov and Solovtsov \cite{SS1,S1}. The second term
in (\ref{VDR}), which  violates the DR, corresponds to the
contribution to the integral (\ref{cauchy}) coming from the
``Landau branch cut''. It is represented by the contour integral
\begin{equation}
\label{L-part} d_{\rm L}(Q^{2})=-{1\over
2\pi\imath}\oint_{C_{L}^{+}}{d_{\rm
RGI}(\zeta)\over{\zeta-Q^{2}}}d\,\zeta,
\end{equation}
taken round the circle $\{\zeta: \zeta=s_{\rm L}+s_{\rm
L}\exp{(\imath\phi)},-\pi<\phi\leq \pi\}$ in the positive
(anti-clockwise) direction.

The perturbation theory approximation to the hadronic spectral
function is calculated by inserting the series (\ref{RGI_PT}) into
the inversion formula (\ref{spectral}). An important point is that
the ``Landau part'' $d_{\rm L}(Q^{2})$ does not contribute into
the spectral function, provided that $s>0$. To see this, let us
evaluate this contribution to the spectral function, with the aid
of formula (\ref{L-part}),
\begin{equation}
\begin{array}{l}
{\displaystyle 2v_{1}(s)|_{\rm L}={1\over
2\pi\imath}\oint_{-s-\imath 0}^{-s+\imath 0}{d_{\rm L}(z)\over
z}d\,z}={\displaystyle-\left({1\over
2\pi\imath}\right)^{2}\oint_{-s-\imath 0}^{-s+\imath 0}{d\,z\over
z}\oint_{C_{L}^{+}} {d_{\rm RGI}(\zeta)\over {\zeta-z}}d\,\zeta=}\\
\qquad {}-{\displaystyle {1\over 2\pi\imath}\oint_{C_{\rm L}^{+}}
d_{\rm RGI}(\zeta)\left\{{1\over 2\pi\imath}\oint_{-s-\imath
0}^{-s+\imath 0}{1\over z(\zeta-z)}d\,z\right\}}d\,\zeta,
\end{array}
\end{equation}
here we have interchanged the order of integration in the repeated
integral. Let us consider the integral under  braces. For
$\zeta\neq 0$ the integrand has two simple poles inside the
contour of integration. It follows from the theorem of residues
that this integral vanishes, provided $s>0$,
$$
{1\over 2\pi\imath}\oint_{-s-\imath 0}^{-s+\imath 0}{1\over
z(\zeta-z)}d\,z\equiv 0,
$$
and the same result holds for $\zeta=0$. We have thus found that
only the ``analytic component'' $d_{\rm APT}(Q^{2})$ gives a
finite contribution into the hadronic spectral function. Using DR
(\ref{DR})  and  inversion formula (\ref{spectral}), one finds the
expression for the spectral function in terms of the effective
spectral density
\begin{equation}
\label{relation} v_{1}^{\rm pQCD}(s)\equiv v_{1}^{\rm APT}(s)
={1\over 2}(1+r(s)),
\end{equation}
where
\begin{equation}
\label{rs0} r(s)={1\over \pi}\int_{s}^{\infty}{\rho_{\rm
eff}(\sigma)\over\sigma}d\,\sigma.
\end{equation}
Note that formulas (\ref{relation}) and (\ref{rs0}) were
previously obtained in the context of APT (see \cite{Y} and
\cite{MSa}). With the help of formula (\ref{relation}),   we
express the ``perturbative component" of the total ``experimental"
Adler function in terms of the effective spectral density
\begin{equation}
\label{tmp2} D_{\rm pQCD}(Q^{2},s_{p})=\int_{s_p}^{\infty}{\cal
K}(Q^{2},s)(1+r(s))d\,s
\end{equation}
where  we have introduced the notation  ${\cal
K}(Q^{2},s)=Q^{2}/(s+Q^{2})^{2}$. Integrating (\ref{tmp2}) by
parts we obtain a more convenient representation
\begin{equation}
\label{practical} D_{\rm
pQCD}(Q^{2},s_{p})={Q^{2}\over{s_{p}+Q^{2}}}(1+r(s_{p}))
-{Q^{2}\over\pi}\int_{s_{p}}^{\infty}{\rho_{\rm
eff}(\sigma)\over{\sigma(\sigma+Q^{2})}}d\,\sigma.
\end{equation}
Let us now evaluate  power suppressed corrections to the total
``experimental'' Adler function (\ref{Exp}). We  may rewrite the
perturbative component of the Adler function  identically
\begin{equation}
\label{tmp3}
\begin{array}{l}
\displaystyle{D_{\rm pQCD}(Q^{2},s_{p})=}D_{\rm
APT}(Q^{2})-2\int_{0}^{s_{p}}{\cal K}(Q^{2},s)v_{1}^{\rm APT}(s)d\,s\\
\displaystyle{ \qquad{}\qquad{}=D_{\rm RGI}(Q^{2})-d_{\rm
L}(Q^{2}) -2\int_{0}^{s_{p}}{\cal K}(Q^{2},s)v_{1}^{\rm
APT}(s)d\,s},
\end{array}
\end{equation}
in the first line of (\ref{tmp3}) we have used the definition of
the analytic image of the Adler function
\begin{equation}
D_{\rm APT}(Q^{2})=1+d_{\rm APT}(Q^{2})=2\int_{0}^{\infty}{\cal
K}(Q^{2},s)v_{1}^{\rm APT}(s)d\,s,
\end{equation}
which is easily deduced from the discussion given above. The last
equality on the right of (\ref{tmp3}) follows from formula
(\ref{VDR}). The power suppressed part of the total
``experimental'' Adler function is determined as
\begin{equation}
\label{PS} D_{\rm p.s.}(Q^{2},s_{\rm p})=D_{\rm
``exp''}(Q^{2})-D_{\rm RGI}(Q^{2}).
\end{equation}
Combining formulas (\ref{Exp}), (\ref{VDR}) and (\ref{tmp3}), we
rewrite formula (\ref{PS}) in the form
\begin{equation}
\begin{array}{l}
D_{\rm p.s.}(Q^{2},s_{\rm p})=
D_{\rm exp}(Q^{2},s_{\rm p})+D_{\rm pQCD}(Q^{2},s_{\rm p})-D_{\rm RGI}(Q^{2})\\
\qquad{}=\int_{0}^{s_{\rm p}}K(Q^{2},s)2 v_{1}^{\rm
exp}(s)d\,s-d_{\rm L}(Q^{2}) -\int_{0}^{s_{\rm
p}}K(Q^{2},s)2v_{1}^{\rm APT}(s)d\,s \label{tmp4}.
\end{array}
\end{equation}
From  definitions (\ref{L-part}) and (\ref{tmp2}), we  obtain the
asymptotic formulas, for $Q^{2}\rightarrow \infty$,
\begin{equation}
\label{AE} {\cal K}(Q^{2},s)\approx Q^{-2}+{\cal O}(sQ^{-4}),\quad
d_{\rm L}(Q^{2})\approx c_{\rm L}\Lambda^{2}Q^{-2}+{\cal
O}(\Lambda^{4}Q^{-4}),
\end{equation}
where $\Lambda$ denotes the conventional ${\overline {\rm
MS}}$-scheme QCD parameter ($\Lambda\equiv\Lambda_{\overline {\rm
MS}}$). Since the parameter $s_{\rm L}$ is proportional to
$\Lambda^{2}$ \footnote{The expressions for $s_{\rm L}$ in terms
of $\Lambda$ up to fourth order in perturbation theory may be
found in \cite{my3} (see, also, Appendix A).}, the coefficient
$c_{\rm L}$ is a positive number independent of $\Lambda$
\begin{equation}
\label{cL} c_{\rm L}=\Lambda^{-2}{1\over 2\pi\imath }\oint_{\rm
C_L^{+}}d_{\rm RGI}(\zeta)d\,\zeta={1\over 2\pi}{s_{\rm L}\over
{\Lambda}^{2}}\int_{-\pi}^{\pi}d_{\rm RGI}(s_{\rm L}+s_{\rm
L}e^{\imath \phi})d\,\phi.
\end{equation}
Using formulas (\ref{tmp4}) and (\ref{AE}), we write asymptotic
expansion for $D_{\rm p.s.}(Q^{2},s_p)$. It follows from the OPE
that the leading term in the asymptotic expansion, proportional to
$Q^{-2}$, vanishes if the quarks are massless. This leads to the
equation \footnote{The FOPT version of this equation reads
$\int_{0}^{s_{\rm p}}v_{1}^{\rm FO}(s)d\,s=\int_{0}^{s_{\rm p}}
v_{1}^{\rm exp}(s)d\,s$ (see \cite{PPR}). }
\begin{equation}
\label{tmp5} c_{\rm L}\Lambda^{2}+\int_{0}^{s_{\rm p}}2v_{1}^{\rm
APT}(s)d\,s\equiv c_{\rm L}\Lambda^{2}+s_{\rm p}+\int_{0}^{s_{\rm
p}}r(s)d\,s =\int_{0}^{s_{\rm p}} 2v_{1}^{\rm exp}(s)d\,s,
\end{equation}
the first equality in Eq.~(\ref{tmp5}) follows from the relation
(\ref{relation}).  Using Eq.~(\ref{rs0}), by partial integration
we find
\begin{equation}
\label{tmp8} \int_{0}^{s_{\rm p}}r(s)d\,s=s_{\rm p}r(s_{\rm
p})+{1\over \pi}\int_{0}^{s_{\rm p}}\rho_{\rm
eff}(\sigma)d\,\sigma,
\end{equation}
here we have   used  the relation $sr(s)\rightarrow 0$ as
$s\rightarrow 0$, which holds in every order of perturbation
theory. Combining Eqs.~(\ref{tmp5}) and  (\ref{tmp8}), we obtain
\begin{equation}
\label{tmp6} c_{\rm L}\Lambda^{2}+s_{\rm p}(1+r(s_{\rm
p}))+{1\over\pi}\int_{0}^{s_{\rm p}}\rho_{\rm
eff}(\sigma)d\,\sigma =\int_{0}^{s_{\rm p}}2v_{1}^{\rm
exp}(s)d\,s.
\end{equation}
The coefficient $c_{\rm L}$ is calculated numerically from formula
(\ref{cL}). In this calculations we use the exact (explicit)
two-loop running coupling and  exact (numeric) four-loop running
coupling \footnote{Application of the explicit series solution
(\ref{sersol}) for the four-loop coupling yield the same results.
}. In particular,  using the next-to-next-to-leading order
approximation ($\rm N^{2}LO$) to the Adler function constructed in
terms of the exact two-loop order running coupling we find
\begin{equation}
c_{\rm L}|_{\rm two-loop\, \rm beta}=0.421163.
\end{equation}
The numerical values of the coefficient $c_{\rm L}$ evaluated in
the $\overline  {\rm MS}$ scheme   in the case of the four-loop
order exact (numeric) running coupling are listed in Table
\ref{tab:30}. In the calculations we have used the approximations
to the Adler function of increasing order \footnote{we will use
the abbreviation ${\rm N}^{k}{\rm LO}$ to denote the order ${\cal
O}(\alpha_{s}^{k+1})$ approximation to the Adler function. }. For
the unknown ${\cal O}(\alpha_{s}^{5})$ correction to the Adler
function, we employ the geometric estimate
$d_{5}=d_{4}(d_{4}/d_{3})=378$ \cite{DHZ}.

\begin{table}
\caption{The numerical values of the coefficient $c_{\rm L}$ in
the $\overline{\rm MS}$ scheme as calculated from formula
(\ref{cL}) in the case of the four-loop order $\beta$-function.}
\label{tab:30}
\begin{tabular}{lccccc}\hline\noalign{\smallskip}
&\multicolumn{5}{c}{Approximations to the Adler
function}\\\cline{2-6}\noalign{\smallskip} & $\rm LO$& $\rm NLO$ &
$\rm N^{2}LO$ &$\rm N^{3}LO$&$\rm
N^{4}LO$\\\hline\noalign{\smallskip}
$c_L$& 0.301262& 0.453421& 0.555401& 0.651373& 0.721687\\
\hline\noalign{\smallskip}
\end{tabular}
\end{table}

It follows from the mixed representation (\ref{GD}) for the
spectral function that one may calculate in perturbation theory
the decay rate of the $\tau$ lepton into hadrons of invariant mass
larger than $\sqrt{s_{\rm p}}$
\begin{equation}
\label{rate} R^{\rm pert.}_{\tau,V}|_{s>s_{\rm p}}=6|V_{\rm
ud}|^{2}S_{\rm EW}\int_{s_{\rm
p}}^{m_{\tau}^{2}}w_{\tau}(s)v_{1}^{\rm APT}(s)d\,s,
\end{equation}
where
$$ w_{\tau}(s)={1\over m_{\tau}^{2}}\left(1-{s\over
m_{\tau}^{2}}\right)^{2}\left(1+2{s\over m_{\tau}^{2}}\right),
$$
$V_{\rm ud}$  and $S_{\rm EW}$ denote the flavor mixing matrix
element and an electro-weak correction term respectively
\cite{BNP}. Equation (\ref{rate}) reduces to
\begin{equation}
\label{second} \int_{s_{\rm
p}}^{m_{\tau}^{2}}w_{\tau}(s)v_{1}^{\rm APT}(s)d\,s =\int_{s_{\rm
p}}^{m_{\tau}^{2}}w_{\tau}(s)v_{1}^{\rm exp}(s)d\,s.
\end{equation}
Using  relation (\ref{relation}), we  express the left hand side
of (\ref{second}) in terms of the effective spectral density. By
integrating by parts, after some algebra, we obtain
\begin{equation}
\label{tmp7}
\begin{array}{l}
\displaystyle{\int_{s_{\rm p}}^{m_{\tau}^{2}}w_{\tau}(s)v_{1}^{\rm
APT}(s) d\,s={1\over 4}\left(1-{s_{\rm p}\over m_{\tau}^{2}
}\right)^{3}\left(1+{s_{\rm p}\over m_{\tau}^{2}
}\right)(1+r(s_{\rm p}))}\\\qquad{}\displaystyle{-{1\over
4\pi}\int_{s_{\rm p}}^{m_{\tau}^{2}}}{\rho_{\rm eff}(s)\over
s}\left(1-{s\over m_{\tau}^{2}}\right)^{3}\left(1+{s\over
m_{\tau}^{2}}\right)d\,s.
\end{array}
\end{equation}

To clarify the difference between the APT and $\rm APT^{+}$
frameworks, a few comments are in order: i) In APT the spectral
function is determined through formula (\ref{relation}) in the
entire region $0<s<\infty$, whereas in $\rm APT^{+}$ this formula
holds only for $s>s_{\rm p}$. ii) In APT the RG improved
approximation to the Adler function, $D_{\rm RGI}(Q^2)$, is
replaced  with corresponding ``analytic'' image $D_{\rm APT}(Q^2)$
from the outset. Then formula (\ref{PS}), the definition of the
power suppressed contributions to the Adler function must be
suitably modified (see \cite{cvetic}). In this paper, we do not
mention this procedure. iii) To parametrize our results we use the
standard coupling constant $\alpha_{s}(m_{\tau}^{2})$, whereas in
APT the results are parametrized in terms of the analytic coupling
$\alpha_{s}(m_{\tau}^{2})_{\rm an}$

\section{Numerical Results for the Parameters}
\label{sec:3} To extract the parameters $s_{\rm p}$ and $\Lambda$
from the data we have to solve the system of equations
\begin{eqnarray}
\label{transc1}
\Phi_{1}(s_{\rm p},{\Lambda}^{2})&=&\int_{0}^{s_{\rm p}}v_{1}^{\rm exp}(s)d\,s,\\
\label{transc2} \Phi_{2}(s_{\rm p},{\Lambda}^{2})&=&\int_{s_{\rm
p}}^{m_{\tau}^{2}}w_{\tau}(s)v_{1}^{\rm exp}(s)d\,s,
\end{eqnarray}
where the functions $\Phi_{1,2}$ are defined as
\begin{equation}
\label{Phi1} \Phi_{1}(s_{\rm p},{\Lambda}^{2})={s_{\rm p}\over
2}(1+r(s_{\rm p}))+{1\over 2\pi}\int_{0}^{s_{\rm p}}\rho_{\rm
eff}(\sigma)d\,\sigma+{c_{\rm L}\over2}\Lambda^{2},
\end{equation}
\begin{equation}
\label{Phi2}
\begin{array}{l}
 \displaystyle{\Phi_{2}(s_{\rm p},{\Lambda}^{2})=(1-{\hat s}_{\rm
 p})^{3}(1+{\hat
s}_{\rm p}){(1+r(s_{\rm p}))\over
4}}\\
\qquad {}{}\displaystyle{-{1\over 4\pi}\int_{{\hat s}_{\rm
p}}^{1}{\rho_{\rm eff}(m_{\tau}^{2}y)\over y}(1-y)^{3}(1+y)d\,s},
\end{array}\end{equation}
with ${\hat s}_{\rm p}=s_{\rm p}/m_{\tau}^{2}$. The right hand
sides of Eqs.~(\ref{transc1})-(\ref{transc2}) are determined in
terms of the empirical function $v_{1}^{\rm exp}(s)$. We
reconstruct the experimental vector spectral function  from the
ALEPH 2005 spectral data for the vector invariant  mass squared
distribution which is publicly available \cite{compilation} (see
Appendix B). The spectral function is measured at discrete points
of the energy squared. To interpolate the spectral function
between these points we use cubic splines.

We solve the system of equations (\ref{transc1})-(\ref{transc2})
numerically     using various approximations to the Adler
function. Since the system  is transcendental it has more than one
solution. In Table \ref{tab:1}, we present the first reasonable
solution for the parameters obtained at next-to-next-to-leading
order (${\rm{N^{2}}}\rm{LO}$). From the Table, we see that the
predictions for $s_{\rm p}$ are stable with respect to the loop
corrections to the $\beta$-function. In this regard, the
predictions for the QCD scale parameter is more sensitive. The two
values of $\Lambda$ obtained with the two- and four-loop
$\beta$-functions differ in about 10\%. However, this corresponds
to the small difference $\alpha_{s}(m_{\tau}^{2})|_{\rm
four-loop}-\alpha_{s}(m_{\tau}^{2})|_{\rm two-loop}\approx
0.0017$.

The solution for $s_{\rm p}$ obtained with the two-loop running
coupling should be compared  with the    estimate $s_{\rm
p}=1.60\pm 0.17$ extracted in \cite{PPR} from the earlier ALEPH
data. Our prediction for the central value of $s_{\rm p}$ (see
Table \ref{tab:1}) is greater in about 7\%. However, with the more
accurate data, we have obtained smaller experimental errors on the
parameters (see Appendix B).  Our estimate for the central value,
$\Lambda|_{{\rm \{two-loop\,\beta\}}}=383 \,{\rm MeV}$ is somewhat
above the value $\Lambda|_{{\rm \{two-loop\,\beta\}}}=372\,{\rm
MeV}$ accepted in \cite{PPR}. However, one should keep in mind
that in \cite{PPR} only one equation, the FOPT counterpart of
Eq.~(\ref{transc1}), has been utilized.

Note that the system (\ref{transc1})-(\ref{transc2}) permits one
more solution for the parameters in the range $200\,{\rm
MeV}<\Lambda<600\,{\rm MeV}$ (see Table \ref{tab:2}). An
attractive feature of this solution is that it predicts a smaller
value for the onset of perturbation theory: $s_{\rm p}=0.607\,{\rm
GeV}^{2}\approx m_{\rho}^{2}$ ($m_{\rho}$ stands for the
$\rho$-meson mass). However, considering  current  status of
$\alpha_s$ we find the  extracted value for the strong coupling
constant too large. For this reason, we decline this solution.

We also  determine the experimental uncertainties on the
parameters coming from the uncertainties of the vector invariant
mass squared distribution. The correlations between the errors of
the distribution are properly  taken into account. Cumbersome
technical details of the error analysis are relegated into
Appendix B.

\begin{table}
\caption{The first solution for the parameters $s_{\rm p}$ and
$\Lambda=\Lambda_{\overline{\rm MS}}$   obtained at ${\rm
N}^{2}{\rm LO}$. The  two- and four-loop  running couplings have
been used. The extracted values of the strong coupling constant
$\alpha_{s}(m_{\tau}^{2})$ are also given. The error bars refer to
the experimental uncertainty only.} \label{tab:1}
\begin{tabular}{lcc}\hline\noalign{\smallskip}
Observable&\multicolumn{2}{c}{Approximation to the
$\beta$-function}\\\cline{2-3}\noalign{\smallskip} & Two-loop&
Four-loop\\\hline\noalign{\smallskip}

$s_{\rm p}$\,\,${\rm GeV}^{2}$ &$1.711\pm 0.054$& $1.709\pm 0.054$\\
$\Lambda\,\,{\rm GeV}$&$0.383\pm 0.034$&$0.348\pm 0.030$\\
$\alpha_{s}(m_{\tau}^{2})$&$0.320\pm 0.015$&$0.321\pm
0.016$\\\hline\noalign{\smallskip}
\end{tabular}
\end{table}

\begin{table}
\caption{The same as in Table \ref{tab:1}  for the case of the
second solution for the parameters.}\label{tab:2}
\begin{tabular}{lcc}\hline\noalign{\smallskip}
Observable&\multicolumn{2}{c}{Approximation to the
$\beta$-function}\\\cline{2-3}\noalign{\smallskip} & Two-loop&
Four-loop\\\hline\noalign{\smallskip}

$s_{\rm p}$\,\,${\rm GeV}^{2}$ &$0.606\pm 0.003$& $0.607\pm 0.003$\\
$\Lambda\,\,{\rm GeV}$&$0.583\pm 0.018$&$0.522\pm 0.016$\\
$\alpha_{s}(m_{\tau}^{2})$&$0.417\pm 0.010$&$0.424\pm
0.011$\\\hline\noalign{\smallskip}
\end{tabular}
\end{table}

It is useful to  determine the so-called  indicative estimates of
the theoretical uncertainties  on the numerical values of the
parameters (for the definition see \cite{KKP}). This requires us
to test convergence of the numerical results order-by order in
perturbation theory. We use consecutive approximations to the
Adler function  from $\rm{LO}$ to $\rm{N}^{4}\rm{LO}$. For the
unknown ${\cal O}(\alpha_{s}^{5})$ correction, we use the
geometric estimate $d_{5}=d_{4}(d_{4}/d_{3})=378\pm 378$
\cite{DHZ}. The results for the extracted values of the parameters
are presented in Table~\ref{tab:3}.
\begin{table}
\caption{Numerical values for the  parameters in the
$\overline{\rm MS}$ scheme extracted from the $\tau$ data
order-by-order within the modified procedure based on ${\rm
APT}^{+}$.} \label{tab:3}

\begin{tabular}{lccccc}
\hline\noalign{\smallskip}
Observable&\multicolumn{5}{c}{Approximation to the Adler
function}\\\cline{2-6}\noalign{\smallskip}&$\rm{LO}$ &$\rm{NLO}$&
$\rm{N^{2}LO}$& $\rm{N}^{3}\rm{LO}$&$\rm{N}^{4}\rm{LO}$
\\\noalign{\smallskip}\hline\noalign{\smallskip}
$s_{\rm p}$\,\,${\rm GeV}^{2}$&1.707&1.710&1.709&1.707&1.705\\
$\Lambda$\,\,${\rm GeV}$ & 0.486& 0.378& 0.348 & 0.332 & 0.323  \\
$\alpha_{s}(m_{\tau}^{2})$ &0.401 &0.337&0.321&0.313 &0.308
\\\noalign{\smallskip}\hline
\end{tabular}
\end{table}
Formally, we may write a series for the numerical value of the
coupling constant as follows
$$
\alpha_{s}(m_{\tau}^{2})|_{\rm{N}^{4}\rm{LO}}=\alpha_{s}(m_{\tau}^{2})|_{\rm{LO}}+\sum_{k=1}^{4}\Delta_{k},
$$
where
$\Delta_{k}=\alpha_{s}(m_{\tau}^{2})|_{\rm{N}^{k}\rm{LO}}-\alpha_{s}(m_{\tau}^{2})|_{\rm{N}^{k-1}\rm{LO}}$.
Using the numbers listed   in Table \ref{tab:3}  (we use
abbreviation $\rm{APT}^{+}$ for the modified APT accepted in this
paper) we obtain the series
\begin{equation}
\label{numseries1}
\alpha_{s}(m_{\tau}^{2})|_{\rm{N}^{4}\rm{LO}}^{{\rm
APT}^{+}}=0.401-0.064-0.016-0.009-0.005.
\end{equation}
The changes of the leading term induced by the consecutive
corrections in the series are found to be: 15.9\%, 4.0\%, 2.2\%
 and 1.2\%. It is interesting to compare the
series (\ref{numseries1}) with its counterpart obtained within
standard CIPT. Using the  standard CIPT to analyze the same data
(for details see Appendix C) we obtain the series
\begin{equation}
\label{numseries2}
\alpha_{s}(m_{\tau}^{2})|_{\rm{N}^{4}\rm{LO}}^{\rm
CIPT}=0.485-0.095-0.023-0.013-0.007.
\end{equation}
We see that within CIPT  the corrections provide slightly larger
changes of the leading term: 19.6\%, 4.7\%, 2.7\% and 1.4\%. One
finds that $\Delta_{k}(\rm{CIPT})/\Delta_{k}(\rm{APT^{+}})\approx
1.2$ for $k=1-4$ .  So that the series (\ref{numseries1})
converges slightly rapidly than the series (\ref{numseries2}). The
indicative estimate of the theoretical uncertainty is determined
as a half of the last retained term in the series \cite{KKP}
\footnote{In \cite{KKP} this definition of the uncertainty has
been used within FOPT. }. As pointed out in \cite{KKP},  the error
defined in this way is heuristic and indicative. The actual values
of the theoretical errors related to the uncalculated higher order
terms in  the perturbation theory series for the decay rate might
be even larger (see, for example, papers
\cite{BNP,KKP,KAST,ALEPH1}). In this paper, however, we shall
consider only  the indicative theoretical errors. From the series
(\ref{numseries1}), we obtain the estimates
\begin{eqnarray}
\alpha_{s}(m_{\tau}^{2})|_{\rm{NLO}}&=&0.337\pm 0.016_{\rm exp}\pm
0.032_{\rm th}\nonumber \\
\alpha_{s}(m_{\tau}^{2})|_{\rm{N}^{2}\rm{LO}}&=& 0.321\pm
0.016_{\rm exp}  \pm 0.008_{\rm th}\nonumber\\
\alpha_{s}(m_{\tau}^{2})|_{\rm{N}^{3}\rm{LO}}&=& 0.313\pm
 0.014_{\rm exp}\pm 0.004_{\rm th}\nonumber\\
 \alpha_{s}(m_{\tau}^{2})|_{\rm{N}^{4}\rm{LO}}&=&0.308\pm
0.014_{\rm exp}\pm 0.002_{\rm th},
 \label{ierrors1}
\end{eqnarray}
here we have also included the experimental errors. Analogically,
from the CIPT series (\ref{numseries2}), one obtains
\begin{eqnarray}
\alpha_{s}(m_{\tau}^{2})|_{\rm{NLO}}&=&0.390 \pm 0.011_{\rm
exp}\pm
0.048_{\rm th}\nonumber \\
\alpha_{s}(m_{\tau}^{2})|_{\rm{N}^{2}\rm{LO}}&=& 0.367 \pm
0.009_{\rm exp}\pm 0.012_{\rm th}\nonumber\\
 \alpha_{s}(m_{\tau}^{2})|_{\rm{N}^{3}\rm{LO}}&=&0.354\pm
 0.008_{\rm exp}\pm
 0.007_{\rm th}\nonumber\\
 \label{ierrors2}
\alpha_{s}(m_{\tau}^{2})|_{\rm{N}^{4}\rm{LO}}&=&0.347\pm
0.008_{\rm exp}\pm 0.003_{\rm th},
\end{eqnarray}
 The $\rm{N}^{4}\rm{LO}$ estimates in (\ref{ierrors1}) and
(\ref{ierrors2}) correspond to the  central value $d_5=378$.  The
additional theoretical error in the coupling constant induced from
the uncertainty in the fifth order unknown coefficient  $(d_5=378
\pm 378)$ takes the values $0.0045$ $(\approx 1.5\%)$ and $0.0065$
$(\approx 1.9\%)$ in the new  and standard extraction procedures
respectively \footnote{With $d_{5}=756$, we have obtained
$\alpha_{s}(m_{\tau}^{2})|_{\rm APT^{+}}=0.3035$ and
$\alpha_{s}(m_{\tau}^{2})|_{\rm CIPT}=0.3407$. }. Comparing the
numbers in  formulas (\ref{ierrors1}) and (\ref{ierrors2}), we see
that the indicative estimates of the theoretical error are smaller
within the new procedure. In contrast to this,    the experimental
errors on the values of $\alpha_s$ increases by the factor of 1.76
within the new procedure  . It is remarkable that a more reliable
estimate of the theoretical error presented in \cite{DHZ} is close
to the $\rm N^{3}LO$ and $\rm N^{4}LO$ values of the indicative
error given in formula (\ref{ierrors2}).

Similarly, determining the indicative theoretical errors on the
parameter $s_{\rm p}$, we find the stable results
\begin{eqnarray}
s_{\rm p}|_{\rm NLO}&=&1.710 \pm 0.054_{\rm exp}\pm 0.002_{\rm th}\quad{\rm GeV^{2}} \nonumber \\
s_{\rm p}|_{\rm{N}^{2}\rm{LO}}&=&1.709\pm 0.054_{\rm exp}\pm 0.001_{\rm th}\quad{\rm GeV^{2}} \nonumber\\
s_{\rm p}|_{\rm{N}^{3}\rm{LO}}&=&1.707\pm 0.054_{\rm exp}\pm
0.001_{\rm th}\quad{\rm GeV^{2}}
 \nonumber\\
 \label{ierrors3}
s_{\rm p}|_{\rm{N}^{4}\rm{LO}}&=&1.705\pm 0.054_{\rm exp}\pm
0.001_{\rm th}\quad{\rm GeV^{2}}.
\end{eqnarray}

Notice that the ratio $\alpha_s(s_{\rm
p})/\alpha_s(m_{\tau}^{2})\approx 1.22$ is not  large. However,
the $\rm APT^{+}$ expansion formally depends on the small energy
scale $\sqrt{ s_{\rm p}}\approx 1.31 \rm GeV$. So, it is
reasonable to justify the applicability of the perturbation theory
in the $\rm APT^{+}$ framework. The issue of the applicability of
perturbation theory in $\tau$ decays has been previously addressed
in \cite{GN}.   It was pointed out \cite{GN} that this question is
phenomenological one, and it cannot be answered yet from
theoretical grounds. In particular, the decay rate of the $\tau$
lepton into hadrons of invariant mass squared smaller than $s_{0}
$ ($s_{0}<m_{\tau}^{2}$) has been analyzed within FOPT. Using the
ALEPH spectral data, the authors of \cite{GN} have deduced that
the rate can be calculated in pQCD with high accuracy for
$s_0>s_{\rm min.}=0.7\,\rm GeV^{2}$. Note that our estimate for
$s_{\rm p}$ clearly satisfies this condition, $s_{\rm p}/s_{\rm
min.}\approx 2.4$. Nevertheless, it is desirable to investigate
numerically the convergence of the perturbative expansion  within
$\rm APT^{+}$.  Let us derive the expansion for the $\tau$-lepton
decay rate from formula (\ref{rate}). The integral on the right of
(\ref{rate}) can be approximated by a non-power series. To derive
the non-power series, we express the spectral function in terms of
the effective spectral density using formulas (\ref{relation}) and
(\ref{rs0}). Then we expand the function $\rho_{\rm eff}(s)$ in
perturbation theory using formulas (\ref{RGI_PT}) and
(\ref{specfun}). So, we obtain
\begin{equation}
\label{rate3} \hat{R}_{\tau,\rm V}^{\rm pert.}|_{s>s_{\rm
p}}=R^{\rm pert.}_{\tau,\rm V}|_{s>s_{\rm p}}/\{6|V_{\rm
ud}|S_{\rm EW}\}=\sum_{k=0}^{5}d_{k}{\mathfrak
{A}}_{k}(m_{\tau}^{2},s_{\rm p})
\end{equation}
where
\begin{eqnarray}
{\mathfrak A}_{0}(m_{\tau}^{2},s_{\rm p}) &=&f(s_{\rm p}/m_{\tau}^{2}) ,\\
{\mathfrak A}_{k\geq 1}(m_{\tau}^{2},s_{\rm p}) &=& r_{k}(s_{\rm
p})f(s_{\rm p}/m_{\tau}^{2})-{1\over\pi}\int_{s_{\rm
p}}^{m_{\tau}^{2}}{f(\sigma/m_{\tau}^{2})\over
\sigma}\rho_{k}(\sigma)d\,\sigma,
\end{eqnarray}
here we have used the notations: $f(x)={1\over 4}(1-x)^{3}(1+x)$
and
\begin{eqnarray}
\label{notat}
\rho_{k}(\sigma)&=& \rm{Im}\{a_{s}^{k}(-\sigma-\imath 0)\},\nonumber\\
r_{k}(s_{\rm p})&=&{1\over \pi}\int_{s_{\rm
p}}^{\infty}{\rho_{k}(\sigma)\over\sigma}d\,\sigma.
\end{eqnarray}
The first term  in the series (\ref{rate3}), ${\mathfrak A}_{0}$,
corresponds to the (modified) parton level contribution to the
rate. We calculate the functions ${\mathfrak A}_{k}$  numerically
by using analytic expressions for the functions $\rho_{k}(\sigma)$
(see formula (\ref{effsp}) in Appendix A). In the calculation, we
employ the four-loop running coupling.  For the parameters $s_{\rm
p}$ and $\Lambda\equiv \Lambda_{\overline{\rm MS}}$, we use the
numerical values from the Table \ref{tab:3}, namely,  the values
extracted from the ALEPH data within $\rm APT^{+}$ at $\rm
N^{4}LO$.  Using analytically known coefficients $d_{k}$, $k=0-4$
and the estimate $d_5=378$, we obtain from Eq.~(\ref{rate3}) the
expansion
\begin{equation}
\label{expan1}
\begin{array}{l} \hat{R}_{\tau,\rm V}^{\rm pert.}|_{s>s_{\rm
p}}=0.3747\cdot 10^{-1}+0.3275\cdot 10^{-2}+0.3937\cdot 10^{-3}+0.9270\cdot 10^{-4}\\
\qquad{}+0.3304\cdot 10^{-4}+(0.6047\cdot 10^{-5})\approx 0.04127.
\end{array}
\end{equation}
Consider now the non-power expansion for the perturbation theory
correction $\delta^{(0)}$ obtained within CIPT  \cite{DP}  (see
Appendix C)
\begin{equation}
\label{CINE} \delta^{(0)}_{\rm CI}=\sum_{k=1}d_{k}{\cal
A}_{k}(m_{\tau}^{2}),
\end{equation}
where
$$
{\cal A}_{k}(m_{\tau}^{2})={1\over\pi}\int_{0}^{\pi}{\rm
Re}\{(1-e^{\imath\phi})(1+e^{\imath\phi})^{3}a_{s}^{k}(m_{\tau}^{2}e^{\imath\phi})\}d\,\phi,
$$
to calculate these functions numerically, we employ for the  scale
parameter $\Lambda$ the numerical value   extracted from the ALEPH
data within CIPT at $\rm N^{4}LO$ (see Table \ref{tab:11}). At
$\rm N^{4}LO$, the expansion (\ref{CINE}) can be rewritten as
\begin{equation}
\label{expan2}
\delta^{(0)}_{\rm CI}=0.1513+0.3081\cdot
10^{-1}+0.1276\cdot 10^{-1}+0.9012\cdot 10^{-2}+(0.5233\cdot
10^{-2})\approx 0.2091.
\end{equation}
Comparing the numerical expansions in Eqs.~(\ref{expan1}) and
(\ref{expan2}), one sees that the $\rm APT^{+}$ series
(\ref{expan1}) displays a faster convergence. In the CIPT
expansion (\ref{expan2}), the corrections provide a 38\% change of
the leading term. In contrast, in the $\rm APT^{+}$ expansion
(\ref{expan1}) the corrections provide only a 16\% change of the
leading term (we recall that the leading QCD correction in
(\ref{expan1}) is the second term in the series).

The rapid convergence of the series (\ref{expan1}) may be
explained due to the specific properties of the expansion
functions ${\mathfrak {A}}_{k}(m_{\tau}^{2},s_{\rm p})$. The set
of functions $\{{\mathfrak {A}}_{k}(m_{\tau}^{2},s_{\rm p})\}$ can
be viewed as a generalization of the analogical set of functions
considered in the Shirkov-Solovtsov APT (for properties of the APT
expansion functions see \cite{Shirk}). In Table \ref{tab:12}, we
have compared functions ${\mathfrak {A}}_{k}(m_{\tau}^{2},s_{\rm
p})$ with the functions $ {\cal A}_{k}(m_{\tau}^{2})$. For the
sake of comparison, we also include in the Table  the powers of
the ``couplant" $a_{s}(m_{\tau}^{2})$. It is seen from the Table
that the functions ${\mathfrak {A}}_{k}(m_{\tau}^{2},s_{\rm p})$
decrease with k much more rapidly than the functions ${ \cal
A}_{k}(m_{\tau}^{2})$ and $a_{s}^{k}(m_{\tau}^{2})$.

\begin{table}
\caption{Comparison of the  expansion functions ${\mathfrak
{A}}_{k}(m_{\tau}^{2},s_{\rm p})$, ${\cal A}_{k}(m_{\tau}^{2})$
and the powers of the ``couplant" $a_{s}(m_{\tau}^{2})$. The
four-loop ``couplant"  is calculated using the value
$\Lambda=0.3225\,\rm GeV$. To calculate the functions ${\mathfrak
{A}}_{k}(m_{\tau}^{2},s_{\rm p})$, we have used the values
$\Lambda|_{\rm N^{4}LO}=0.3225\,\rm GeV$ and $s_{\rm p}|_{\rm
N^{4}LO}=1.7053\,\rm GeV$ obtained within $\rm APT^{+}$. To
calculate the functions ${\cal A}_{k}(m_{\tau}^{2})$  we have used
the value $\Lambda|_{\rm N^{4}LO}=0.395\,\rm GeV$ obtained from
the ALEPH data within CIPT.} \label{tab:12}

\begin{tabular}{llll}
\hline\noalign{\smallskip} k&$a_{s}^{k}(m_{\tau}^{2})$&${\cal
A}_{k}(m_{\tau}^{2})$&${\mathfrak
{A}}_{k}(m_{\tau}^{2},s_{\rm p})$\\

\noalign{\smallskip}\hline\noalign{\smallskip}
1&$0.9797\cdot 10^{-1}$&0.1511&$0.3275\cdot 10^{-2}$\\
2&$0.9599\cdot 10^{-2}$ &$0.1876\cdot 10^{-1}$&$0.2400\cdot 10^{-3}$\\
3&$0.9405\cdot 10^{-3}$&$0.2000\cdot 10^{-2}$&$0.1455\cdot 10^{-4}$\\
4&$0.9214\cdot 10^{-4}$&$0.1834\cdot 10^{-3}$&$0.6733\cdot 10^{-6}$\\
5&$0.9028\cdot 10^{-5}$&$0.1383\cdot 10^{-4}$&$0.1599\cdot 10^{-7}$\\
\noalign{\smallskip}\hline
\end{tabular}
\end{table}

Usually, it is convenient to perform evolution of the $\alpha_ s$
results to the reference scale $M_z=91.187\,\rm GeV$. This is done
by using RG equation and appropriate matching conditions at the
heavy quark (charm and bottom) thresholds (see \cite{rodrigo1} and
literature therein). The three-loop level matching conditions in
the $\overline{\rm MS}$ scheme  were derived in \cite{chetyrk1}.
In this paper, we  follow the work \cite{rodrigo2}, where a very
accurate analytic approximation to the four-loop running coupling
was suggested. We perform the matching at the matching scale
$m_{\rm th}=2\mu_{\rm h}$ where $\mu_{\rm h}$ is a scale invariant
$\overline{\rm MS}$ mass of the heavy quark $\mu_{\rm h}=
{\overline m}_{\rm h}(\mu_{\rm h})$. We assume for the scale
invariant $\overline{\rm MS}$ masses the values $\mu_{\rm
c}=1.27^{+0.07}_{-0.11}\,\rm GeV$ and $\mu_{\rm
b}=4.20^{+0.17}_{-0.07}\,\rm GeV$ \cite{PPRD}. Following
\cite{rodrigo2}, we evaluate   the central value and error of
$\alpha_{s}(M_{z}^{2})$ according to the formulas
$$\alpha_{s}(M_{z}^{2})=(\alpha_{s}^{+}(M_{z}^{2})+\alpha_{s}^{-}(M_{z}^{2}))/2
\quad{}{\rm and}\quad{}
\Delta\alpha_{s}(M_{z}^{2})=(\alpha_{s}^{+}(M_{z}^{2})-\alpha_{s}^{-}(M_{z}^{2}))/2$$
where $\alpha_{s}^{\pm}(M_{z}^{2})$ denote the values obtained
from $\alpha_s^{\pm}(m_{\tau}^{2})=\alpha_s(m_{\tau}^{2})\pm\Delta
\alpha_s(m_{\tau}^{2})$. In the evolution procedure, we have used
the exact numeric four-loop running coupling   \footnote{ We have
confirmed that   the approximate analytical coupling derived in
\cite{rodrigo2}   leads practically  to the same numerical
results.}. In Table~\ref{tab:4}, we compare the estimates for
$\alpha_{s}(M_{z}^{2})$ obtained within the new (${\rm APT}^{+}$)
and standard (CIPT)   procedures.
\begin{table}
\caption{Estimates for $\alpha_{s}(M_{z}^{2})$ obtained  from the
ALEPH $\tau$ lepton  decay vector data order-by-order in
perturbation theory. The results obtained within    $\rm APT^{+}$
and  $\rm CIPT$ are compared. Two errors are given, the
experimental (first number) and the error from the evolution
procedure (second number). } \label{tab:4}

\begin{tabular}{lcc}
\hline\noalign{\smallskip}
 Perturbative order&$\alpha_{s}(M_{z}^{2})|_{\rm{APT}^{+}}$&
$\alpha_{s}(M_{z}^{2})|_{\rm{CIPT}}$\\

\noalign{\smallskip}\hline\noalign{\smallskip}
 $\rm{N}^{2}\rm{LO}$&$0.1187\pm
0.0019\pm 0.0005$&$0.1238\pm
0.0009\pm 0.0005$\\
$\rm{N}^{3}\rm{LO}$&$0.1176\pm 0.0018\pm 0.0005$&$0.1224\pm
0.0009\pm 0.0005$\\
 $\rm{N}^{4}\rm{LO}$&$0.1170\pm
0.0018\pm 0.0005$&$0.1217\pm
0.0009\pm 0.0005$\\

\noalign{\smallskip}\hline
\end{tabular}
\end{table}

Finally, for the sake of comparison, let us   extract the
numerical values for the coupling constant  from $\tau$ decay
(V+A) data using the renormalization scheme invariant extraction
procedure (RSI) of \cite{KKP}. This procedure is based on FOPT.
For the experimental value of the perturbative part of the $\tau$
decay rate in the non-strange (V+A) channel, we assume the updated
value, presented in \cite{BJ},
$$
\delta^{(0)}_{\rm exp}|_{V+A}=0.2042\pm 0.0050_{\rm exp}.
$$
For consistency reasons we use the $\overline{\rm MS}$ scheme
$\beta$-function to the k-loop order with the $\rm N^{k-1}LO$
approximation to the Adler function. In Table (\ref{tab:21}) we
compare numerical values for $\alpha_{s}(m_{\tau}^{2})$ obtained
within the two approaches, RSI and $\rm APT^{+}$. The relevant
channels which have been used to extract  the coupling are
indicated by subscripts. One sees from the Table, that beyond $\rm
NLO$ there is a good agreement between the two methods of the
$\alpha_{s}(m_{\tau}^{2})$ determination.

\begin{table}
\caption{Comparison of the RSI and $\rm APT^{+}$ determinations of
the $\overline{\rm MS}$ coupling constant from the $\tau$-decay
data. Experimental errors are  given only.}

\label{tab:21}
\begin{tabular}{|l|l|l|}\hline
Perturbative order& $\alpha_{s}(m_{\tau}^{2})|_{V+A}^{\rm RSI}$&$\alpha_{s}(m_{\tau}^{2})|^{\rm APT^{+}}_{V}$\\
&&\\\hline

$\rm NLO$ &$0.278\pm 0.003$ &$0.335\pm 0.016$ \\
  $\rm N^{2}LO$ &$0.319\pm 0.004$ & $0.321\pm 0.016$\\
$\rm N^{3}LO$& $0.312\pm 0.004$ & $0.313\pm 0.014$\\\hline
\end{tabular}
\end{table}

\section{Numerical Results for the ``Experimental''Adler Function}
\label{sec:4} Looking at the numbers in  Table \ref{tab:1}, we see
that our estimates for the parameters are somewhat different  than
those  used previously in \cite{PPR}. Hence, it is sensible to
recalculate the experimental Adler function in the infrared
region. Another reason to do this is the appearance of the
improved $\tau$ data \cite{compilation}. More importantly, it is
desirable to carry out the error analysis too. Furthermore, in
contrast to \cite{PPR}, in our calculations we will employ
$\rm{APT}^{+}$.

The ``experimental'' Adler function and its QCD component are
tabulated in Table~\ref{tab:5}. The QCD component of the
``experimental'' Adler function is calculated numerically at $\rm
N^{2}LO$ from formula (\ref{practical}). In the calculations we
employ the four-loop running coupling. For the parameters $s_{\rm
p}$ and $\Lambda$, we  use the values from Table~\ref{tab:1}. The
absolute $(\pm 1\sigma)$ and relative (in percents) experimental
errors  of the ``experimental'' Adler function are also tabulated.
The  error analysis is described in Appendix B. We see  from the
Table  that the pQCD component has sizeable contribution to the
total ``experimental" Adler function. This contribution increases
monotonically with $Q$ from 10\% (at $Q=0.1\,\,{\rm GeV}$) to 36\%
(at $Q=1\,\,{\rm GeV}$).
\begin{table}
\caption{Comparison of the ``experimental'' Adler function $D_{\rm
``exp"}(Q^{2})$ with its QCD component $D_{\rm pQCD}(Q^{2},s_{\rm
p})$ at low momenta. The perturbative component is evaluated
within $\rm APT^{+}$ at $\rm N^{2}LO$ using the four-loop running
coupling. The absolute and relative statistical errors of the
``experimental" Adler function are tabulated.} \label{tab:5}
\begin{tabular}{ccccc}
\hline\noalign{\smallskip}
 $Q\,\,{\rm GeV}$& $D_{\rm ``exp"}(Q^{2})$ & $D_{\rm pQCD}(Q^{2},s_{\rm p})$  &
$\sigma(D_{\rm ``exp"})$& rel.err. \\

\noalign{\smallskip}\hline\noalign{\smallskip}

0.1&0.0649& 0.0063    &0.0061  &9.5\%  \\
0.2&0.2300& 0.0249     &0.0198 &8.6\%  \\
0.3&0.4354& 0.0545     &0.0333 &7.7\%  \\
0.4&0.6320& 0.0933     &0.0426 &6.7\%  \\
0.5&0.7944& 0.1391     &0.0473 &6.0\%  \\
0.6&0.9162& 0.1895     &0.0484 &5.3\%  \\
0.7&1.0016& 0.2426     &0.0471&4.7\%   \\
0.8&1.0583& 0.2965     &0.0445&4.2\%   \\
0.9&1.0942& 0.3497     &0.0412&3.8\%   \\
1.0&1.1157& 0.4013     &0.0377&3.4\%   \\
\noalign{\smallskip}\hline
\end{tabular}
\end{table}

To test the stability of the numerical results with regards  to
the higher order  corrections to the $\beta$-function, we have
compared two results for the ``experimental" Adler function that
are obtained with the two- and four-loop exact running couplings.
The pQCD component of the Adler function has been evaluated within
$\rm{APT}^{+}$ at ${\rm N^{2}LO}$. For the parameters $s_{\rm p}$
and $\Lambda$, we have used the central values given in Table
\ref{tab:1}.   In the region $Q=0-1.5\,\,{\rm GeV}$, the
difference between using the two- or four-loop approximation to
the $\beta$-function is found to be quite small ($\sim 0.05$\%).
The approximation corresponding to the two-loop running coupling
takes slightly large values.

To test the stability of   numerical results with regards   to
higher order corrections to the Adler function, we use various
approximations to the pQCD component (see Table~\ref{tab:6}). We
see from the Table that the differences between the consecutive
approximations to the ``experimental'' Adler function slowly
increase as a function of the scale. Already, the leading order
approximation provides a very accurate result. At $Q=1.5
\,\,\rm{GeV}$ (where the changes induced by the loop correction
take maximal values) the differences between consecutive
approximations (i.e. the differences between the $\rm N^{k-1}LO$
and $\rm N^{k}LO$ approximations) take the values 0.05\%, 0.03\%,
0.03\% and 0.03\% for $k=1, 2, 3$ and 4 respectively.
\begin{table}
\caption{Different approximations to the ``experimental'' Adler
function as a function of the scale. The function $D^{(k)}_{\rm
``exp"}(Q^{2})$ has the pQCD component evaluated within ${\rm
APT}^{+}$ at ${\rm N}^{(k-1)}{\rm LO}$. To construct this
component, we employ the four-loop order running coupling.}
\label{tab:6}
\begin{tabular}{llllll}
\hline\noalign{\smallskip}

$Q\,\,{\rm GeV}$&$D_{\rm ``exp"}^{(1)}(Q^{2})$ &$D_{\rm
``exp"}^{(2)}(Q^{2})$ &$D_{\rm ``exp"}^{(3)}(Q^{2})$&
$D_{\rm ``exp"}^{(4)}(Q^{2})$&$D_{\rm ``exp"}^{(5)}(Q^{2})$\\

\noalign{\smallskip}\hline\noalign{\smallskip}
0.1&0.06494 &0.06494 &0.06494 &0.06494 &0.06494 \\
0.2&0.23003 &0.23005 &0.23004 &0.23003 &0.23002 \\
0.3&0.43541 &0.43546 &0.43544 &0.43541 &0.43539 \\
0.4&0.63196 &0.63205 &0.63201 &0.63196 &0.63192 \\
0.5&0.79431 &0.79444 &0.79438 &0.79430 &0.79424 \\
0.6&0.91613 &0.91631 &0.91623 &0.91612 &0.91604 \\
0.7&1.0015  &1.0017  &1.0016  &1.0015  &1.0014 \\
0.8&1.0582  &1.0585  &1.0583  &1.0582  &1.0580 \\
0.9&1.0940  &1.0943  &1.0942  &1.0940  &1.0938 \\
1.0&1.1154  &1.1158  &1.1157  &1.1154  &1.1152 \\
1.5&1.1321  &1.1327  &1.1324  &1.1320  &1.1317 \\
\noalign{\smallskip}\hline
\end{tabular}
\end{table}

It is instructive to investigate numerically the convergence
property of the non-power  series for the perturbation theory
component $D_{\rm pQCD}(Q^{2},s_{\rm p})$. The non-power  series
is obtained from formula (\ref{practical}) by using perturbation
theory expansions for the function $\rho_{\rm eff}(s)$ and $r(s)$
(see formulas (\ref{RGI_PT}), (\ref{specfun}) and (\ref{rs0})).
The non-power series read
\begin{equation}
\label{nonpower} D_{\rm pQCD}(Q^{2},s_{\rm
p})=\sum_{k=0}d_{k}{\mathfrak D}_{k}(Q^{2},s_{\rm p}),
\end{equation}
where
\begin{eqnarray}
{\mathfrak D}_{0}(Q^{2},s_{\rm p}) &=& {Q^{2}\over(s_{\rm p}+Q^{2})},\\
{\mathfrak D}_{k\geq 1}(Q^{2},s_{\rm p}) &=& {Q^{2}\over(s_{\rm
p}+Q^{2})}r_{k}(s_{\rm p})-{Q^{2}\over\pi}\int_{s_{\rm
p}}^{\infty}{\rho_{k}(\sigma)\over \sigma(\sigma+Q^{2})}d\,\sigma,
\end{eqnarray}
the functions $\rho_{k}(\sigma)$ and $r_{k}(s_{\rm p})$ are
defined in Eqs.~(\ref{notat}). Let us truncate the non-power
expansion (\ref{nonpower})  at ${\rm N}^{4}\rm{LO}$ (i.e. for
$k=5$). Using the ${\rm N}^{4}\rm{LO}$ estimates for the
parameters given in Table \ref{tab:3}, we evaluate the ratios of
the consecutive terms of the series
$${\cal R}_{k}(Q^{2})=(d_{k}/d_{k-1}){\mathfrak D}_{k}(Q^{2},s_{\rm
p})/{\mathfrak D}_{k-1}(Q^{2},s_{\rm p}),$$ for $k=1-5$. In Table
\ref{tab:7}, we tabulate numerical values of these ratios in the
region $Q=0.1-1.5\,\,{\rm GeV}$.
\begin{table}
\caption{The ratios of the consecutive terms in the non-power
series (\ref{nonpower}) as a function of the scale.}\label{tab:7}
\begin{tabular}{llllll}
\hline\noalign{\smallskip} $Q\,\,{\rm GeV}$& ${\cal R}_{1}(Q^{2})$
& ${\cal R}_{2}(Q^{2})$&${\cal R}_{3}(Q^{2})$&
${\cal R}_{4}(Q^{2})$&${\cal R}_{5}(Q^{2})$\\
\noalign{\smallskip}\hline\noalign{\smallskip}
0.1& 0.077  & 0.112 & 0.228 & 0.366& 0.238 \\
0.3& 0.077  & 0.111 & 0.228 & 0.366& 0.240\\
0.5& 0.077  & 0.111 & 0.227 & 0.366& 0.242\\
0.7& 0.076  & 0.110 & 0.226 & 0.367& 0.245\\
1.0& 0.075  & 0.109 & 0.225 & 0.367& 0.251\\
1.3& 0.073  & 0.108 & 0.223 & 0.368& 0.257\\
1.5& 0.072  & 0.107 & 0.222 & 0.368& 0.261\\
\noalign{\smallskip}\hline
\end{tabular}
\end{table}
It is seen from the Table, that the magnitudes of the  ratios are
sufficiently small  to guarantee  fast numerical convergence of
the series: ${\cal R}_{k}(Q^{2})\leq 0.368$  ($k=1-5$) for all
values of $Q$ in the considered interval.

Let us now compare our numerical results on  the ``experimental"
Adler function with the previous results  of work \cite{PPR}.
First, we repeat the calculation within the approach of \cite{PPR}
using the improved data. Assuming the value
$\Lambda_{\overline{\rm MS}}=372\pm 76\,\rm MeV$ used in
\cite{PPR}, we solve numerically the FOPT counterpart of the
equation (\ref{transc1}).   Thus we find the solution $s_{\rm
p}=1.621\pm0.163\,\,\rm GeV^{2}$. The central value of this
estimate is slightly large, by  0.021, than the value obtained in
\cite{PPR}. Using the values $s_{\rm p}=1.621\,\,\rm GeV^{2}$ and
$\Lambda_{\overline{\rm MS}}=372\,\,{\rm MeV}$, we  calculate the
``experimental" Adler function within the (modified) FOPT  at
${\rm N}^{2}\rm LO$.  This should be compared with the new
approximation  computed in the same order within ${\rm APT}^{+}$.
In the case of  ${\rm APT}^{+}$,  we use the values
$\Lambda_{\overline{\rm MS}}=383\,{\rm MeV}$ and $s_{\rm
p}=1.711\,\rm GeV^{2}$  (see Table \ref{tab:1}).  To be consistent
with \cite{PPR},  we use the two-loop exact running coupling. In
Table \ref{tab:8}, we compare numerically two approximations  to
the ``experimental'' Adler function, the functions $D_{\rm
``exp"}^{\rm FOPT}(Q^{2})$ and $D_{\rm ``exp"}^{\rm
APT^{+}}(Q^{2})$.
\begin{table}
\caption{Comparison of the ``experimental'' Adler functions
evaluated within the modified FOPT and $\rm APT^{+}$. The pQCD
components of the functions are constructed at  $\rm N^{2}LO$ with
the two-loop order running coupling. The pQCD component of
$D(Q^{2})|^{\rm FOPT}_{\rm ``exp"}$ corresponds to the values
$\Lambda_{\overline{\rm MS}}=372\,\rm MeV$ and $s_{\rm
p}=1.621\,\rm GeV^{2}$. The pQCD component of  $ D(Q^{2})|^{\rm
APT^{+}}_{\rm ``exp"}$ corresponds to the values
$\Lambda_{\overline{\rm MS}}=383\,\rm MeV$ and $s_{\rm
p}=1.711\,\rm GeV^{2}$. The relative difference between these
functions is also tabulated.}\label{tab:8}
\begin{tabular}{llll}
\hline\noalign{\smallskip} $Q\,\,{\rm GeV}$& $D(Q^{2})|^{\rm
FOPT}_{\rm ``exp"}$ &$D(Q^{2})|_{\rm ``exp"}^{\rm APT^{+}}$ &
 rel.diff.\\
\noalign{\smallskip}\hline\noalign{\smallskip}

0.1& 0.0651 & 0.0649  & 0.31\%\\
0.2& 0.2305  & 0.2301   & 0.20\%\\
0.3& 0.4364  & 0.4355   & 0.21\% \\
0.4& 0.6336  & 0.6321   & 0.24\%\\
0.5& 0.7967  & 0.7945   & 0.28\%\\
0.6& 0.9191  & 0.9164   & 0.29\%\\
0.7& 1.0050  & 1.0018   & 0.32\%\\
0.8& 1.0621  & 1.0586   & 0.33\%\\
0.9& 1.0982  & 1.0945   & 0.34\%\\
1.0& 1.1197  & 1.1160   & 0.33\%\\
1.1& 1.1316  &1.1280   & 0.32\%\\
1.2& 1.1371  &1.1337    &0.30\%\\
1.3& 1.1386  &1.1355    &0.27\%\\
1.4& 1.1377  &1.1350    &0.24\%\\
1.5& 1.1354  &1.1330    &0.21\%\\

\noalign{\smallskip}\hline
\end{tabular}
\end{table}
From the Table, we see that the functions are  close, but $D_{\rm
``exp"}^{\rm FOPT}(Q^{2})>D_{\rm ``exp"}^{\rm APT^{+}}(Q^{2})$.
The relative difference between the functions in the considered
region varies in the interval $0.20\%-0.34\%$.

\section{Conclusion}
\label{sec:5} We have extracted the numerical values of the strong
coupling constant $\alpha_s$  and the parameter $s_{\rm p}$ (the
square of the boundary energy) from the non-strange vector $\tau$
data provided by ALEPH.  Based on the semi-empirical
representation (\ref{GD}) for the hadronic non-strange vector
spectral function, we have developed a modified extraction
procedure. This procedure enabled us to avoid direct application
of the  standard OPE formalism in Minkowski space. The
distinguishing feature of our analysis is that  we have determined
the two parameters ($\alpha_{s}$ and $s_{\rm p}$) simultaneously
from the data.

In Sect. 2, we have derived  a violated  DR for the RG improved
perturbation theory correction to the Adler function, the formula
(\ref{VDR}). Using the violated DR, we have shown that the
perturbation theory component of the ``experimental'' spectral
function is determined via the APT formula (\ref{relation}). This
determines    the hadronic spectral function in terms of the
effective spectral density, $\rho_{\rm eff}(\sigma)$, the basic
object of the perturbation theory calculation. We have  obtained a
convenient expression for the pQCD part of the ``experimental"
Adler function in terms of the effective spectral density, the
formula (\ref{practical}). Using the violated DR (\ref{VDR}), we
have determined the power suppressed corrections to the
``experimental" Adler function via  the formula (\ref{tmp4}).
Making further use of the consistency condition from the OPE for
the ``experimental'' Adler function, we have derived
Eq.~(\ref{tmp6}). This equation relates the parameters $s_{\rm p}$
and $\Lambda$ to the values of the hadronic spectral function on
the range $0<s<s_{\rm p}$. Next we used the {\it ansatz}
(\ref{GD})  for the spectral function to calculate the $\tau$
decay rate $R_{\tau,V}|_{s>s_{\rm p}}$. In this way, we have
derived Eq.~(\ref{second}) which relates the parameters to the
integral of the hadronic spectral function (multiplied by known
function) over the range $s_{\rm p}<s<m_{\tau}^{2}$.

In Sect. 3, we have solved, numerically, the obtained system of
equations for the parameters $s_{\rm p}$ and
$\Lambda\equiv\Lambda_{\overline{\rm MS}}$. To examine  the
convergence of the  numerical results for the  parameters, we have
used perturbation theory approximations to the Adler function up
to the   ${\rm N}^{4}{\rm LO}$. The indicative estimates of the
theoretical errors  \cite{KKP} are used as a criterion of the
quality of the approximations. Based on this criterion, we have
demonstrated that the new framework ($\rm APT^{+}$), compared to
the standard one (CIPT), provides a better numerical convergence
for the extracted value of the coupling constant
$\alpha_s(m_{\tau}^{2})$. It is remarkable that  the central
values of the coupling constant extracted within $\rm APT^{+}$ in
different orders of perturbation theory become systematically
smaller as compared to the corresponding values obtained  within
CIPT (cf. formulas (\ref{ierrors1}) and (\ref{ierrors2})). The
changes in the central values are not within the quoted
experimental and theoretical errors. At $\rm N^{3}LO$ The central
values  of $\alpha_s(m_{\tau}^{2})$  in formulas (\ref{ierrors1})
and (\ref{ierrors2})  differ from each other in about 2.7 standard
deviation, if the error is determined within $\rm APT^{+}$,
$\sigma=\sqrt{\sigma_{\rm exp.}^{2}+\sigma_{\rm th.}^{2}}\approx
0.0151$. With the error obtained in CIPT, $\sigma\approx 0.0107$,
one finds even large difference, $3.8\,\sigma$ \footnote{Due to
the larger experimental error obtained within $\rm APT^{+}$,
$\sigma_{\rm APT}/\sigma_{\rm CIPT}\approx 1.4$.}.

We have examined the stability of the numerical  value for the
duality point $s_{\rm d}$ with regard to perturbation theory
corrections.  We have obtained a surprisingly stable result (see
formulas (\ref{ierrors3})) $s_{\rm d}=1.71\pm 0.05_{\exp.}\pm
0.00_{\rm th.}\,\,\rm GeV^{2}$. Our prediction for the central
value of this parameter is higher in about 7\% than the value
presented previously in \cite{PPR}.

Having included into analysis the fourth order coefficient $d_4$,
we achieved excellent agreement between the lattice and tau-decay
determinations of the strong coupling constant (at $\rm N^{4}LO$
the central value of the constant given in Table (\ref{tab:4})
coincides with the central value quoted in (\ref{latt})). For this
reason we believe that $\rm APT^{+}$ provides better approximation
than CIPT.

To justify the applicability of $\rm APT^{+}$ in calculations of
the $\tau$-lepton decay rates, we examine numerically the $\rm
APT^{+}$ series. The $\rm APT^{+}$ expansion for the rate
$R_{\tau,V}|_{s>s_{\rm p}}$ represents asymptotic expansion over a
non-power set of specific functions $\{{\mathfrak
A}_{n}(m_{\tau},s_{\rm p})\}$ rather than the powers of
$a_{s}(m_{\tau}^{2})$.
The $\rm APT^{+}$ and CIPT series for the rates $R_{\tau,\rm
V}|_{s>s_{\rm p}}$ and $R_{\tau,\rm V}$ have been compared
numerically. We have confirmed that the $\rm APT^{+}$ expansion
displays a faster convergence.

Our approach  confirms  that there is a theoretical systematic
uncertainty not included in the error assessments obtained in
previous studies by ignoring the higher order OPE contributions,
the conclusion achieved in work~\cite{Malt}. In this connection,
our study suggests that the truncated OPE series cannot
approximate  sufficiently accurately  the integrals of the
spectral function over the low energy region $0<s<s_p\sim
1.7\,\,{\rm GeV}^{2}$.

 In Sect. 4, we have recalculated the ``experimental'' Adler
function within the ${\rm APT}^{+}$ prescription using the new
estimates  for the parameters  $s_{\rm p}$ and $\Lambda$. In
addition, we have determined the errors on this function coming
from the uncertainties of the parameters and   spectral function.
Numerical results for the Adler function obtained within ${\rm
APT}^{+}$ have been found to be remarkable   stable in
perturbation theory (see Tables \ref{tab:6}  and \ref{tab:7} ).

In Appendix A, we have given practical formulas for numerical
calculation of the ${\overline {\rm MS}}$ running coupling at
higher orders. The  Lambert-W solutions to the RG equation is
reviewed. An  accurate analytic approximation to the effective
spectral density $\rho_{\rm eff}(\sigma)$ at higher orders is
derived. In Appendix B, we have derived formulas (within ${\rm
APT}^{+}$) for calculating  the experimental uncertainties on the
extracted values of the parameters and on the ``experimental''
Adler function.  In Appendix C, we have analyzed the ALEPH
non-strange vector spectral data within the  standard CIPT
prescription. Namely, we have performed some necessary
calculations needed for comparing the CIPT and ${\rm APT}^{+}$
prescriptions (see Sect. \ref{sec:3}).

The procedure suggested here can  be obviously extended to analyze
the non-strange $\tau$-data from the axial-vector (A) and vector
plus axial-vector (V+A) channels.  To check the reliability of the
new extraction procedure, it is desirable to compare the  V, A,
and V+A determinations of the coupling constant. A similar
framework may be constructed on the basis of FOPT. This will
enable us to estimate total theoretical errors on the extracted
values of the parameters $\Lambda$ and $s_{\rm p}$. It should be
remarked that a shortcoming of  the ansatz (\ref{GD}) is that it
completely ignores the non-perturbative contributions to the
spectral function coming from the region $s>s_p$. The importance
of these contributions for accurate determination of the coupling
constant has been demonstrated in recent studies \cite{CGP,Boito}.
We hope to report on these aspects in future publications.

\begin{acknowledgements}
I am very grateful for the support of my colleagues at Department
of Theoretical Physics at Andrea Razmadze Mathematical Institute.
I wish to thank S. Peris and Z. Zhang  for helpful correspondence
and discussions regarding experimental aspects of spectral
function determinations. I thank the referees for valuable
comments and corrections. The present work has been partially
supported by the Georgian National Science Foundation under grants
No GNSF/ST08/4-405 and No GNSF/ST08/4-400.
\end{acknowledgements}

\appendix

\section{A Series Solution to the Renormalization Group Equation}

In our notation the RG equation for the running coupling  reads
\begin{equation}
\label{RGE} {d\over
d\ln{Q^{2}}}a_{s}=\beta(a_{s})=-a_{s}^{2}\sum_{n=0}\beta_{n}a_{s}^{n},
\end{equation}
where $a_{s}\equiv a_{s}(Q^{2})={\alpha_{s}(Q^{2})/\pi}$ with
$\alpha_{s}(Q^{2})$ being the running coupling. In the
$\overline{\rm MS}$ scheme, the $\beta$-function coefficients are
known to four loops \cite{RVL}. For three active quark flavours
the first four coefficients take the values
$$
\beta_{0}=9/4,\quad \beta_{1}=4,\quad
\beta_{2}=10.05990,\quad\beta_{3}=47.22804.
$$
In general, the RG equation (\ref{RGE}), to an arbitrary order in
perturbation theory, can not be solved explicitly for the
coupling. Usually,  the equation is solved in the asymptotical
regime ${Q^{2}\over\Lambda^{2}}\gg 1$. For our purposes the
asymptotic solution is not suitable, since we need an accurate
solution at relatively low energies. One may, of course, solve the
RG equation numerically. However,  it is more convenient to derive
some accurate analytic approximation to the coupling. Fortunately,
it is possible to solve the RG equation explicitly for the
coupling at the two-loop order \cite{ggk,my1}. The explicit
expression for the $\overline{\rm MS}$ scheme running coupling at
the two-loop order reads
\begin{equation}
\label{exact2L}
a_{s}^{(2)}(Q^{2})=-{\beta_{0}\over\beta_{1}}{1\over
1+W_{-1}(\zeta)}:\quad \zeta=-{1\over
eb_{1}}\left({Q^{2}\over\Lambda^{2}}\right)^{-1/b_{1}},
\end{equation}
where $\beta_{0}$ and $\beta_{1}$ are the first two
$\beta$-function coefficients
$$
\beta_{0}={1\over 4}\left(11-{2\over 3}n_{f}\right),\quad
\beta_{1}= {1\over 16}\left(102-{38\over 3}n_{f}\right),
$$
$b_{1}=\beta_{1}/\beta_{0}^{2}$, $\Lambda\equiv
\Lambda_{\overline{\rm MS}}$ and $W_{-1}$ denotes the branch of
the Lambert W function \cite{lamb}.

The coupling to higher orders may be expanded in powers of the
two-loop order coupling \cite{kour,join}
\begin{equation}
\label{sersol1}
a_{s}^{(k>2)}(Q^{2})=\sum_{n=1}^{\infty}c_{n}^{(k)}a_{s}^{(2)n}(Q^{2}),
\end{equation}
The first two coefficients in this series are universal:
$c_{1}^{(k)}=1$ and $c_{2}^{(k)}=0$ (the condition $c_{2}^{(k)}=0$
follows from the conventional definition of the $\Lambda$
parameter). Other coefficients are determined in terms of the
$\beta$-function coefficients. The  four-loop expressions for the
first several coefficients are given by
$$
 c_{3}^{(4)}={\beta_{2}\over\beta_{0}},\quad
 c_{4}^{(4)}={\beta_{3}\over 2\beta_{0}},\quad c_{5}^{(4)}={5\over
 3}\left({\beta_{2}\over\beta_{0}}\right)^{2}-{\beta_{1}\beta_{3}\over
 6\beta_{0}^{2}},\ldots
$$
It was proved in \cite{my3} that the series has a finite radius of
convergence, and the radius is sufficiently large for all $n_{f}$
values of practical interest.  Partial sums of the series
(\ref{sersol1}) provide very accurate approximations to the higher
order coupling in the wide range of $Q^{2}$. In particular, these
approximations may be safely used at low energies. Thus, for
$Q=1\,\,{\rm GeV}$ and $\Lambda=0.347\,\,{\rm GeV}$, the partial
sum with the first twelve   terms reproduce the exact four-loop
coupling with the precision better than 0.02\%. Using the exact
solution (\ref{exact2L}), the analytical structure of the two-loop
coupling in the complex $Q^{2}$-plane has been determined
\cite{ggk,my1}. It was found that the coupling is an analytic
function in the whole complex plane except the cuts running along
the real $Q^{2}$ axis. Besides the physical cut
$\{Q^{2}:-\infty<Q^{2}<0\}$ corresponding to the logarithmic
singularity at $Q^{2}=0$, there is also the ``Landau'' cut
$\{Q^{2}: 0<Q^{2}<Q_L^{2}\}$ corresponding to the Landau
singularity on the positive $Q^{2}$-axis. The Landau singularity
is a second order algebraic branch point located at
$Q_{L}^{2}=b_{1}^{-b_{1}}\Lambda^{2}$ ($b_{1}^{-b_{1}}\approx
1.205$ for $n_{f}=3$).  The relevant branch of the Lambert
function on the complex $Q^{2}$ plane is determined by the
analytical continuation. For the physical vales of $n_{f }$
($0<n_{f}\leq 6$) the relevant branch   is $W_{-1}$ on the
upper-half plane, whereas the branch  is $W_{1}$ on the lower-half
plane. A limiting value of the coupling from above the physical
cut ($Q^{2}=-\sigma+\imath 0$, $\sigma>0$) is then determined by
\cite{my2}
\begin{equation}
a_{s}^{(2)}(-\sigma+\imath 0)=-{\beta_{0}\over\beta_{1}}{1\over
1+W_{-1}(\zeta_{+})}\quad {\rm with} \quad \zeta_{+}={1\over
eb_{1}}\left({\sigma\over\Lambda^{2}}\right)^{-1/b_{1}}\exp\left\{-\imath\pi\left({1\over
b_{1}}-1\right)\right\},
\end{equation}
similarly, one may write
\begin{equation}
\label{downL} a_{s}^{(2)}(-\sigma-\imath
0)=-{\beta_{0}\over\beta_{1}}{1\over 1+W_{1}(\zeta_{-})}\quad {\rm
with} \quad\zeta_{-}={1\over
eb_{1}}\left({\sigma\over\Lambda^{2}}\right)^{-1/b_{1}}\exp\left\{\imath\pi\left({1\over
b_{1}}-1\right)\right\}.
\end{equation}
Note that the limiting values $a_{s}^{(2)}(-\sigma\pm \imath 0)$
satisfy the Schwarz `principle of reflection',\\
$a_{s}^{(2)}(-\sigma-\imath
0)=\overline{a_{s}^{(2)}(-\sigma+\imath 0)}$, provided that W has
{\it{near
conjugate}} symmetry \cite{lamb}:\\
$W_{k}(\overline{z})=\overline{W_{-k}(z)}$.

We may  construct  an analytic  approximation  to the Adler
function in perturbation theory using formula (\ref{sersol1}) for
the four-loop  running coupling. In this approximation, the Adler
function is an analytic function in the cut complex $Q^{2}$ plane.
It has the branch points  at $Q^{2}=0$ and
$Q^{2}=Q^{2}_{L}=b_{1}^{-b_{1}}\Lambda^{2}>0$. The effective
spectral density (\ref{specfun})   associated with the Adler
function  is then readily calculated, leading to the analytic
expression
\begin{equation}
\label{effsp} \rho_{\rm eff}(\sigma)={\rm
Im}\left\{\sum_{n=1}d_{n}\left(\sum_{m=1}^{N}c^{(4)}_{m}a_{s}^{(2)m}(-\sigma-\imath
0)\right)^{n}\right\},
\end{equation}
where $N\geq 3$ and  $a_{s}^{(2)}(-\sigma-\imath 0)$ is determined
in terms of the W function as given in formula (\ref{downL}).
Formula (\ref{effsp}) considerably simplifies   numerical
calculations of integrals of the effective spectral function
(computer algebra system Maple has an arbitrary precision
implementation of all branches of the Lambert function). In  the
most of the calculations, we have used the truncated series
(\ref{sersol1}) for the four-loop running coupling  preserving the
first twelve terms in the series.  Numerical values of the first
twelve coefficients of the series, for $n_{f}=3$ quark flavours,
are tabulated in Table \ref{tab:9}.

\begin{table}
\caption{Numerical values of the first twelve  coefficients in the
expansion (\ref{sersol1}) for the four-loop running coupling at
$n_{f}=3$ quark flavours.}\label{tab:9}
\begin{tabular}{ll||lll}
\hline\noalign{\smallskip}

 n&$c_{n}$&&n&$c_{n}$
\\
\noalign{\smallskip}\hline\noalign{\smallskip}

1&1&&7&392.1241\\
2&0&&8&2413.463\\
3&3863/864&&9&8248.857\\
4&10.49512&&10&31348.18\\
5&27.09804&&11&147697.8\\
6&190.2642&&12&507565.0\\
 \noalign{\smallskip}\hline
\end{tabular}
\end{table}

\section{The Error Analysis}

In this appendix we will evaluate the experimental errors on the
extracted    values of the parameters. We will also determine the
errors on the ``experimental'' Adler function. The main quantity
employed in our analysis is the vector (non-strange) spectral
function $v_{1}(s)$. It is related with the vector invariant mass
squared distribution (the function $sfm2(s)$ in the notations of
\cite{compilation})
\begin{equation}
\label{kappa} v_{1}(s)=\kappa(s)sfm2(s),
\end{equation}
the kinematical factor $\kappa(s)$ is
\begin{equation} \label{IMD}
\kappa(s)={\cal N}{m_{\tau}^{2}\over(6|V_{\rm ud}|^{2}S_{\rm EW})
}\left({B_V\over
B_{e}}\right){1\over(1-s/m_{\tau}^{2})^{2}(1+2s/m_{\tau}^{2})},
\end{equation}
where $|V_{\rm ud}|=0.9746\pm0.0006$ denotes the flavor mixing
matrix elements, the factor $S_{\rm EW}=1.0198\pm0.0006$ is an
electro-weak correction term,
$m_{\tau}=1777.03^{+0.3}_{-0.26}\,MeV$, $B_{V}=(31.82\pm 0.22 )\%$
and $B_{e}=(17.810\pm 0.039)\%$ are the vector and leptonic
branching fractions respectively (in this paper, we assume these
estimates following  \cite{ALEPH}), $\cal N$ is the normalizing
constant
$$
{\cal
N}=\left\{\int_{0}^{m_{\tau}^{2}}sfm2(s)d\,s\right\}^{-1}\approx{1\over
0.794748}.
$$
The quantity $sfm2(s)$ is measured at 140 equidistant  values of
the energy squared variable  starting from
$s_1=0.0125\,\,\rm{GeV}$ with the bin size
$\Delta_{bin}=0.025\,\,\rm{GeV^{2}}$. Note that the factor
$\kappa(s)$ is determined within an accuracy of less than 1\% for
all values of s in the range $s=0-m_{\tau}^{2}$, while the errors
in determination of $sfm2(s)$ are considerably large. Hence, it is
safe to ignore  the uncertainties coming from the factor
$\kappa(s)$. We may then write
\begin{equation}
\label{errv1}
\sigma_{v_{1}}[k]=|\kappa(s_{k})|\sigma_{sfm2}[k],\quad k=1,...140
\end{equation}
where ${\sigma}_{v_{1}}[k]$ and $\sigma_{sfm2}[k]$ stand for the
standard deviations of $v_{1}(s_{k})$ and $sfm2(s_{k})$
respectively,  and $s_k=s_1+(k-1)\Delta_{bin}$ (k=1,2\ldots). By
definition
\begin{equation}
\label{variance}
 {\sigma}^{2}_{v_{1}}[k]= {\bf E}[(v_{1}(s_{k})-\overline{
v_{1}(s_{k})})^{2}],
\end{equation}
etc \footnote{The symbol ${\bf E}[x]$ refers to the mean value of
x.}. Similarly, the covariance matrices of the errors associated
with the quantities $v_{1}(s)$ and $sfm2(s_{k})$ are related by
the formula
\begin{equation}
\label{covariance}
 {\mathbb C}_{ik}=\kappa(s_{i})\kappa(s_{k}){\tilde\mathbb
 C}_{ik},
\end{equation}
where  ${\mathbb C}_{ik}={\rm cov}(v_{1}(s_{i}),v_{1}(s_{k}))$ and
${\tilde{\mathbb C}}_{ik}={\rm cov}(sfm2(s_{i}),sfm2(s_{k}))$. So
that  respective correlation coefficients  coincide
\begin{equation}
\label{corcoefs} {\mathbb R}_{kl}={{\mathbb
C}_{kl}\over{{\sigma}_{v1}[k]{\sigma}_{v1}[l]}}={{\tilde\mathbb
C}_{kl}\over{{\sigma}_{sfm2}[k]{\sigma}_{sfm2}[l]}}.
\end{equation}
In Table \ref{tab:10}, we present a few measured values of
$sfm2(s)$ and $v_{1}(s)$ together with the associated
uncertainties.
\begin{table}
\caption{A few measured values of $sfm2(s)$ and $v_{1}(s)$. The
standard errors for these quantities  are  indicated.}
\label{tab:10}
\begin{tabular}{lllll}
\hline\noalign{\smallskip}
 $s\,GeV^{2}$& $sfm2(s)$
&$\sigma_{\rm sfm2}(s)$ &
 $v_{1}(s)$&$\sigma_{v_1}(s)$\\
\noalign{\smallskip}\hline\noalign{\smallskip}
0.0875&0.004923 &0.001251 &0.006027  &0.001531\\
0.1125&0.022630 &0.003092 &0.027745  &0.003791\\
0.1375&0.037048 &0.004520 &0.045504  &0.005551\\
0.1625&0.056542 &0.005747 &0.069597  &0.007074  \\
0.1875&0.073407 &0.005875 &0.090583  &0.007250\\
0.2125&0.095429 &0.006541 &0.118095  &0.008095\\
0.2375&0.122440  &0.007574 &0.152005  &0.009403\\
\noalign{\smallskip}\hline
\end{tabular}
\end{table}
Our goal is to estimate the uncertainties on the extracted values
of the parameters induced from the experimental uncertainties of
the spectral function.  We start from the system of
(\ref{transc1})-(\ref{transc2}), which we rewrite in the form
\begin{eqnarray}
\label{equ1} \Phi_{1}(x,y)&=&{\cal E}_{1}(x,\{v_{1}\})\\
\label{equ2} \Phi_{2}(x,y)&=&{\cal E}_{2}(x,\{v_{1}\})
\end{eqnarray}
where we have introduced the notations  $x=s_{\rm p}$,
$y=\Lambda^{2}$ and
\begin{eqnarray}
 {\cal E}_{1}(x,\{v_{1}\})&=&\int_{0}^{x}v_{1}(t)d\,t\nonumber\\
 {\cal E}_{2}(x,\{v_{1}\})&=&\int_{x}^{m_{\tau}^{2}}w_{\tau}(t)v_{1}(t)
d\,t\nonumber,
 \end{eqnarray}
 to avoid a cumbersome notation the superscript ``exp.''
in function $v_{1}^{\rm exp.}(s)$ has been omitted. The solution
to the system (\ref{equ1})-(\ref{equ2}) should be considered as a
functional of $v_{1}(s)$. Let a solution for the parameters, for a
given function $v_{1}(s)$, is
\begin{eqnarray}
x&=&\psi_{1}(\{v_{1}\})\\
y&=&\psi_{2}(\{v_{1}\}).
\end{eqnarray}
we may write $v_{1}(s)={\bar v}_{1}(s)+\delta v_{1}(s)$ , where
${\bar v}_{1}(s)$ is the central (average) value and $\delta
v_{1}(s)$ is the deviation. The central  values of the parameters
should be determined by solving the system
(\ref{equ1})-(\ref{equ2}) for $v_{1}(x)={\bar v}_{1}(x)$ (see, for
example, the book \cite{schenck}) i.e.
\begin{eqnarray}
{\bar x}&=&\psi_{1}(\{{\bar v}_{1}\})\\
{\bar y}&=&\psi_{2}(\{{\bar v}_{1}\}).
\end{eqnarray}
Let us expand the functionals ${\cal E}_{1,2}(x,\{v_{1}\})$  in
powers of a small variation $\delta v_{1}(s)$, preserving  the
terms linear in $\delta x$ and $\delta v_{1}(s)$
\begin{eqnarray}
{\cal E}_{1}(x,\{v_{1}\})&=&\Phi_{1}({\bar x},{\bar y})+\delta x
{\bar
v}_{1}({\bar x})+\int_{0}^{{\bar x}}\delta v_{1}(t)d\,t+\ldots\\
{\cal E}_{2}(x,\{v_{1}\})&=&\Phi_{2}({\bar x},{\bar
y})-w_{\tau}({\bar x}){\bar v}_{1}({\bar x})\delta x+\int_{\bar
x}^{{m_{\tau}^{2}}}w_{\tau}(t)\delta v_{1}(t)d\,t+\ldots,
\end{eqnarray}
here use has been made of the equations ${\cal E}_{1,2}({\bar
x},\{\bar{v}_1\})=\Phi_{1,2}({\bar x},{\bar y})$. Insert these
expansions into Eqs.(\ref{equ1})-(\ref{equ2}) and expand the left
hand sides of the equations in powers of $\delta x$ and $\delta y$
$$
\Phi_{1,2}(x,y)=\Phi_{1,2}({\bar x},{\bar y})+{{\partial
\Phi_{1,2}({\bar x},{\bar y})}\over {\partial{\bar x}}}\delta x
 +{{\partial
\Phi_{1,2}({\bar x},{\bar y})}\over {\partial{\bar y}}}\delta
y+\ldots.
$$
Retaining  terms linear in $\delta x$, $\delta y$ and $\delta
v_{1}$, we are led to the following  linear algebraic system of
equations for the variations $\delta x$ and $\delta y$
\begin{eqnarray}
\label{algeb}
A_{1}\delta x+B_{1}\delta y&=&G_{1}\nonumber\\
A_{2}\delta x+B_{2}\delta y&=&G_{2},
\end{eqnarray}
where
$$
A_{1}={{\partial \Phi_{1}({\bar x},{\bar y})}\over{\partial {\bar
x}}}-{\bar v}_{1}({\bar x}),\quad B_{1}={{\partial \Phi_{1}({\bar
x},{\bar y})}\over{\partial {\bar y}}},\quad G_{1}=\int_{0}^{\bar
x}\delta v_{1}(t)d\,t,
$$
$$
A_{2}={{\partial \Phi_{2}({\bar x},{\bar y})}\over{\partial {\bar
x}}}+w_{\tau}({\bar x}){\bar v}_{1}({\bar x}), \quad
B_{2}={{\partial \Phi_{2}({\bar x},{\bar y})}\over{\partial {\bar
y}}},\quad G_{2}=\int_{\bar x}^{m_{\tau}^{2}}w_{\tau}(t)\delta
v_{1}(t)d\,t.
$$
Using the explicit formulas (\ref{Phi1}) and (\ref{Phi2}), after
some algebra,  we obtain
\begin{eqnarray}
{\partial \Phi_{1}(x,y)\over \partial x}&=&{(1+r(x))\over 2}\nonumber\\
{\partial \Phi_{1}(x,y)\over \partial y}&=&{1\over
2\pi}\int_{-\infty}^{\ln(x/y)}
{\tilde\rho}_{\rm eff}(t)e^{t}d\,t+{c_{\rm L}\over 2}\nonumber\\
{\partial \Phi_{2}(x,y)\over \partial x}&=&-{(1+r(x))\over
4m_{\tau}^{2}}P\left({x\over m_{\tau}^{2}}\right)\nonumber\\
{\partial \Phi_{2}(x,y)\over \partial y}&=&{1\over 4\pi
m_{\tau}^{2}}\int_{\ln(x/y)}^{\ln(m_{\tau}^{2}/y)}{\tilde\rho}_{\rm
eff}(t)e^{t}P\left({ye^{t}\over m_{\tau}^{2}}\right),\nonumber
\end{eqnarray} where
${\tilde\rho}_{\rm eff}(t)\equiv\rho_{\rm eff}(\sigma)$ with
$\sigma=\Lambda^{2}\exp(t)$,  $P(z)=2(z-1)^{2}(2z+1)$, and $r(s)$
is defined in (\ref{rs0}). After solving the system (\ref{algeb}),
we take the averages of the deviations squared (the variances)
\begin{eqnarray}
\overline{{(\delta x)}^{2}}&=&(B_{2}^{2}{\overline{
G_{1}^{2}}}+B_{1}^{2}{\overline{ G_{2}^{2}}}-2B_{1}B_{2}{\overline
{G_{1}G_{2}}})/{\cal D}^{2}\nonumber\\
\label{sigmas} \overline{{(\delta y)}^{2}}&=&(A_{1}^{2}{\overline
{G_{2}^{2}}}+A_{2}^{2}{\overline{
G_{1}^{2}}}-2A_{1}A_{2}{\overline {G_{1}G_{2}}})/{\cal D}^{2},
\end{eqnarray}
where ${\cal D}=A_{1}B_{2}-A_{2}B_{1}$, and the overlined symbols
refer to the averages: $\overline{{(\delta x)}^{2}}=E[(x-{\bar
x})^{2}]$ etc.  To calculate the averages $\overline{ G_{1}^{2}}$,
$\overline{G_{2}^{2}}$ and ${\overline {G_{1}G_{2}}}$  we replace
the integrals $G_{1,2}$ by sums  over the equidistant mesh,  using
the trapezoidal rule,
\begin{equation}
\label{trapez}
 G_{1}\approx\Delta\sum_{k=1}^{n_{\rm p}}g_{k}\delta
v_{1}(t_{k}), \quad G_{2}\approx\Delta\sum_{k=n_{\rm
p}}^{n_{\tau}}\eta_{k}\delta v_{1}(t_{k})
\end{equation}
where $n_{\rm p}=1+[(\overline{s_{\rm p}}-s_{1})/\Delta]_{\rm
round}$, $n_{\tau}=1+[(m_{\tau}^{2}-s_{1})/\Delta]_{\rm round}$
\footnote{here the subscript ``round" refers   to the  integer
nearest to the number inside the square bracket.}, $\Delta$
denotes the width of the mesh which is identified with the bin
size in the data. The mesh points in the  sums are determined by
 $t_{k}=t_{1}+(k-1)\Delta$,  $k=1,2\ldots$,  with $t_{1}=0.0125$ and
 $\Delta=0.025$. The numerical coefficients $g_{k}$  take the values
 $g_{k}=1$ for $1<k<n_{\rm p}$ and $g_{1}=g_{n_{\rm p}}=0.5$.  The
 factors $\eta_{k}$ are determined by
$$
\left\{
\begin{array}{rcl}
\eta_{k}&=&w_{\tau}(t_{k})\quad{\rm if}\quad n_{\rm p}<k<n_{\tau}.\\
\eta_{k}&=&0.5w_{\tau}(t_{k})\quad{\rm if}\quad k=n_{\rm p}\quad
{\rm
or}\quad k=n_{\tau}.\\
\end{array}
\right.
$$
Using  formula (\ref{trapez}) and taking into account  the
definitions (\ref{variance}) and (\ref{covariance}), we  calculate
the required averages
\begin{equation}
\label{G1sq}
{\overline{G_{1}^{2}}}=(\Delta)^{2}\left(\sum_{k=1}^{n_{\rm
p}}g_{k}^{2}{{\sigma}^{2}_{v_{1}}[k]} +2\sum_{k=1}^{n_{\rm
p}-1}\sum_{l>k}^{n_{\rm p}}g_{k}g_{l}{\mathbb C}_{kl}\right),
\end{equation}
\begin{equation}
\label{G2sq}
{\overline{G_{2}^{2}}}=(\Delta)^{2}\left(\sum_{k=n_{\rm
p}}^{n_{\tau}} \eta_{k}^{2}{{\sigma}^{2}_{v_{1}}[k]}
+2\sum_{k=n_{\rm
p}}^{n_{\tau}-1}\sum_{l=k+1}^{n_{\tau}}\eta_{k}\eta_{l}{\mathbb
C}_{kl}\right),
\end{equation}
\begin{equation}
\label{G1G2} \overline{G_{1}G_{2}}=(\Delta)^{2}\sum_{k=1}^{n_{\rm
p}} \sum_{l=n_{\rm p}}^{n_{\tau}}g_{k}\eta_{l}{\mathbb C}_{kl}.
\end{equation}
We are now in a position to determine the uncertainties on the
values of the  ``experimental" Adler function.  They are induced
from the errors of the experimental spectral function  and from
the errors on the parameters $\Lambda$ and $s_{\rm p}$. Let us
represent again the ``experimental" Adler function as a sum of the
two terms showing explicitly the dependence of the terms on the
parameters and on the spectral function
\begin{equation}
\label{decomp} D_{\rm ``exp"}(Q^{2},\Lambda^{2},s_{\rm p}:
v_{1})=D_{\rm exp}(Q^{2},s_{\rm p}: v_{1})+D_{\rm
pQCD}(Q^{2},\Lambda^{2},s_{\rm p}),
\end{equation}
the  experimental and  pQCD parts of the function are determined
as
\begin{eqnarray}
\label{part1}
D_{\rm exp}(Q^{2},s_{\rm p}: v_{1})&=&\int_{0}^{s_{\rm p}}{\cal{K}}(Q^{2},t)v_{1}(t)d\,t\\
\label{part2} D_{\rm pQCD}(Q^{2},\Lambda^{2},s_{\rm
p})&=&\int_{s_{\rm p}}^{\infty}{\cal{K}}(Q^{2},t)v_{1}^{\rm
APT}(t)d\,t,
\end{eqnarray}
where ${\cal{K}}(Q^{2},t)=2Q^{2}/(t+Q^{2})^{2}$, $v_{1}(s)$
denotes the spectral function measured on the experiment
($v_{1}(s)\equiv v_{1}^{\rm exp}(s)$) and $v_{1}^{\rm APT}(s)$ is
the approximation to the spectral function evaluated within APT.
Consider small deviations of the spectral function and the
parameters from their mean values
\begin{equation}
v_{1}(t)=\bar{v}_{1}(t)+\delta v_{1}(t),\quad s_{\rm
p}={\bar{s}}_{\rm p}+\delta s_{\rm p}, \quad
\Lambda^{2}=\overline{\Lambda^{2}}+\delta \Lambda^{2},
\end{equation}
the change of the ``experimental" Adler function under these
variations is
\begin{equation}
\label{var1}
 \delta D_{\rm ``exp"}=\delta D_{\rm exp}+\delta D_{\rm pQCD},
\end{equation}
here we have used abbreviations $D_{\rm exp}\equiv D_{\rm
exp}(Q^{2},s_{\rm p}: v_{1})$ etc. The right hand side of
(\ref{var1}) can be evaluated  using  formulas (\ref{part1}) and
(\ref{part2}). Preserving terms linear in the variations $\delta
v_{1}$, $\delta s_{\rm p}$ and $\delta \Lambda^{2}$, we find
\begin{equation} \delta D_{\rm ``exp"}=\delta_{v_{1}}D_{\rm exp}+E_{{\bar
s}_{\rm p}}\delta s_{\rm p}+E_{{\bar\Lambda^{2}}}\delta
\Lambda^{2}
\end{equation}
where
\begin{eqnarray}
\label{vcoef1} \delta_{v_{1}}D_{\rm exp}&=&\int_{0}^{{\bar
s}_{\rm p}}{\cal K}(Q^{2},t)\delta v_{1}(t)d\,t,\\
\label{vcoef2}
 E_{{\bar s}_{\rm p}}&=&{\cal K}(Q^{2},{\bar s}_{\rm p}){\bar v}_{1}({\bar s}_{\rm p})+
 {{\partial D_{\rm pQCD}(Q^{2},{\bar \Lambda^{2}},{\bar s}_{\rm p})}\over {\partial {\bar
 s}_{\rm p}}},\\
 \label{vcoef3}
E_{{\bar\Lambda^{2}}}&=&{{\partial D_{\rm pQCD}(Q^{2},
 {\bar \Lambda^{2}},{\bar s}_{\rm p})}\over
 {\partial {\bar\Lambda^{2}}}}.
\end{eqnarray}
Using the trapezoidal rule, we approximate the integral on the
right side of Eq.~(\ref{vcoef1})
\begin{equation}
\label{trapez-1} \delta_{v_{1}}D_{\rm exp}\approx
G_{3}=\sum_{k=1}^{n_{\rm p}}g_{k}{\cal K}(Q^{2},t_{k})\delta
v_{1}(t_k),
\end{equation}
where the quantities $n_{\rm p}$, $g_k$ and $t_k$ are defined
below  formula (\ref{trapez}). To calculate the partial
derivatives on the right hand sides of (\ref{vcoef2}) and
(\ref{vcoef3}), we use  explicit formula (\ref{practical}) for the
pQCD part of the Adler function.  We then obtain
$$
{{\partial D_{\rm pQCD}(Q^{2},
 \Lambda^{2},s_{\rm p})}\over
 {\partial s_{\rm p}}}=
 -{Q^{2}\over (Q^{2}+s_{\rm p})^{2}}(1+r(s_{\rm p}))
$$
$$
{{\partial D_{\rm pQCD}(Q^{2},
 {\Lambda^{2}},{s}_{\rm p})}\over
 {\partial {\Lambda^{2}}}}={1\over\pi
 Q^{2}}\int_{\ln(s_{\rm p}/\Lambda^{2})}^{\infty}{{\rm e}^{t}{\tilde\rho}_{\rm eff}(t)\over
 \left(1+{\Lambda^{2}\over Q^{2}}{\rm e}^{t}\right)^{2}}d\,t,
$$
where ${\tilde\rho}_{\rm eff}(t)\equiv \rho_{\rm
eff}(\Lambda^2e^{t}) $ and  to derive the last formula we have
used the relation
$$
{{\partial r(s_{\rm p})}\over{\partial
\Lambda^{2}}}={1\over\pi\Lambda^{2}}\rho_{\rm eff}(s_{\rm p}),
$$
which can be easily  derived  from the definition (\ref{rs0}).
The  mean squared deviation of the ``experimental'' Adler
function  is then determined as a sum of the six averages
\begin{equation}
\label{msqd}
\begin{array}{l}
{\overline {(\delta D_{\rm ``exp"})^{2}}}={\overline
{({\delta}_{v_{1}} D_{\rm exp})^{2}}}+E_{{\bar s}_{\rm
p}}^{2}{\overline{(\delta s_{\rm
p})^{2}}}+E_{{\bar\Lambda^{2}}}^{2}{\overline{(\delta
\Lambda^{2})^{2}}}\\
\qquad+2E_{{\bar s}_{\rm
p}}E_{\overline{\Lambda^{2}}}{\overline{{\delta s_{\rm p}}{\delta
\Lambda^{2}}}}+2E_{{\bar s}_{\rm p}}{\overline{({\delta}_{v_{1}}
D_{\rm exp})\delta s_{\rm
p}}}+2E_{\overline{\Lambda^{2}}}\overline{(\delta_{v_{1}} D_{\rm
exp}){\delta \Lambda^{2}}}.
\end{array}
\end{equation}
With the aid of formula (\ref{trapez-1}),   the first term on the
right of Eq.~(\ref{msqd}) can easily be expressed in terms of the
errors $\sigma_{v_{1}}$ and the covariance matrix ${\mathbb
C}_{kl}$
\begin{equation}
{\overline {({\delta}_{v_{1}} D_{\rm
exp})^{2}}}=\Delta^{2}\left\{\sum_{k=1}^{n_{\rm p}}g_{k}^{2}{\cal
K}^{2}(Q^{2},t_{k})\sigma^{2}_{v1}[k]+2\sum_{k=1}^{ n_{\rm
p}-1}\sum_{l=k+1}^{ n_{\rm p}}g_{k}g_{l}{\cal K}(Q^{2},t_{k}){\cal
K}(Q^{2},t_{l}){\mathbb C}_{kl}\right\}.
\end{equation}
The second and third terms on the right of (\ref{msqd}) are
determined in terms of the errors $\sigma_{s_{\rm p}}$ and
$\sigma_{\Lambda^{2}}$ which we have already evaluated above (see
(\ref{sigmas})). In order to evaluate last three terms on the
right of (\ref{msqd}), we use explicit  expressions for the
deviations $\delta s_{\rm p}$ and $\delta \Lambda^{2}$
\begin{eqnarray}
\label{algsols}
\delta s_{\rm p}={\cal D}^{-1}(B_{2}G_{1}-B_{1}G_{2})\nonumber\\
\delta \Lambda^{2}={\cal D}^{-1}(A_{1}G_{2}-A_{2}G_{1}),
\end{eqnarray}
the  solution to the system (\ref{algeb}). This enable us to write
\begin{equation}
\label{fourth} {\overline{{\delta s_{\rm p}}{\delta
\Lambda^{2}}}}={\cal D}^{-2}\{{(B_{2}A_{1}+B_{1}A_{2}){\overline
{G_{1}G_{2}}}}-B_{2}A_{2}{\overline{G_{1}^{2}}}-B_{1}A_{1}{\overline{G_{2}^{2}}}\},
\end{equation}
the averages on the right hand side of (\ref{fourth}) have been
evaluated above (see Eqs. (\ref{G1sq}), (\ref{G2sq}) and
(\ref{G1G2})). It remains to calculate  the last two averages on
the right hand side of (\ref{msqd}).  Using formulas
(\ref{algsols}) we find
\begin{eqnarray}
\label{fifth} {\overline{({\delta}_{v_{1}} D_{\rm exp})\delta
s_{\rm p}}}=\overline{G_{3}\delta s_{\rm p}}={\cal
D}^{-1}(B_{2}{\overline{G_{1}G_{3}}}-B_{1}{\overline{G_{2}G_{3}}})\\
\label{sixth} \overline{(\delta_{v_{1}} D_{\rm exp}){\delta
\Lambda^{2}}}=\overline{G_{3}\delta \Lambda^{2}}={\cal
D}^{-1}(A_{1}{\overline{G_{2}G_{3}}}-A_{2}{\overline{G_{1}G_{3}}}),
\end{eqnarray}
employing  now the trapezoidal sums  (\ref{trapez}) and
(\ref{trapez-1}), we determine the averages
${\overline{G_{1}G_{3}}}$ and ${\overline{G_{2}G_{3}}}$ in terms
of the correlation coefficients ${\mathbb R}_{k,l}$
\begin{eqnarray}
{\overline{G_{1}G_{3}}}=\Delta^{2}\sum_{k=1}^{{ n}_{\rm
p}}g_{k}\sigma_{v_1}[k]\sum_{l=1}^{{n}_{\rm p}}g_{l}{\cal
K}(Q^{2},t_{l}){\mathbb
R}_{k,l}\sigma_{v_1}[l]\nonumber\\
{\overline{G_{2}G_{3}}}=\Delta^{2}\sum_{k={n}_{\rm p}}^{{
n}_{\tau}}\eta_{k}\sigma_{v_{1}}[k]\sum_{l=1}^{n_{\rm
p}}g_{l}{\cal K}(Q^{2},t_{l}){\mathbb
R}_{k,l}\sigma_{v_{1}}[l]\nonumber.
\end{eqnarray}

\section{Standard CIPT Consideration}

It is instructive to compare  the modified procedure for
extracting the coupling constant  with  the standard procedure
formulated  within conventional CIPT in the $\overline{\rm MS}$
scheme.  The $\tau$ decay rate to the non-strange hadrons in the
vector channel is given by \cite{BNP}
\begin{equation}
\label{rate1} R_{\tau,V}={3\over 2}|V_{\rm ud}|^{2}S_{\rm
EW}(1+\delta_{\rm QCD}+{\delta}_{\rm EW})
\end{equation}
where $\delta_{\rm QCD}$ represents the  QCD corrections, $|V_{\rm
ud}|=0.9746\pm 0.0006$ is the flavor mixing matrix element,
$S_{\rm EW}=1.0198$ is an electro-weak correction term and
${\delta}_{\rm EW}\approx 0.001$ is an additive electroweak
correction (for these values see \cite{ALEPH}). The QCD
contribution is the sum
\begin{equation}
\label{dQCD} \delta_{\rm
QCD}=\delta^{(0)}+\delta^{(2)}+\delta_{\rm NP},
\end{equation}
where $\delta^{(0)}$ is the purely  perturbative contribution,
$\delta^{(2)}$ is the dimension $D=2$ effects from light quark
masses, and $\delta_{\rm NP}$ is the total non-perturbative
contribution: $\delta_{\rm
NP}=\delta^{(4)}+\delta^{(6)}+\delta^{(8)}$ ( $\delta^{(D)}$  are
the OPE terms in powers of $m_{\tau}^{-D}$). We  will use the
estimates  $\delta^{(2)}=(-3.3\pm 3)\times 10^{-4}$ and
$\delta_{\rm NP}=0.0199\pm 0.0027$,   the ALEPH results obtained
within the CIPT approach   \cite{ALEPH}.
 The experimental result   for $\delta^{(0)}$ can be
 determined from the experimental spectral function via the relation
 \begin{equation}
\label{rate2} 1+\delta^{(0)}_{\rm exp}+\delta^{(2)}+\delta_{\rm
NP}+{\delta}_{\rm EW}=4J_{\tau,V}^{\rm exp},
\end{equation}
where
\begin{equation}
\label{Atau} J_{\tau,V}^{\rm
exp.}=\int_{0}^{m_{\tau}^{2}}w_{\tau}(s)v_{1}^{\rm exp}(s)d\,s,
\end{equation}
and explicit expression of the function $w_{\tau}(s)$ is given in
(\ref{rate}). The relation (\ref{rate2}) follows from formulas
(\ref{rate}) and (\ref{rate1}). Let us now determine
 the experimental error on $J_{\tau,V}^{\rm exp}$ induced from the experimental errors  on
$v_{1}^{\rm exp}(s)$. Using the trapezoidal rule, we replace the
integral on the right side of Eq.~(\ref{Atau}) by the sum
\begin{equation}
\label{trpz} J_{\tau,V}^{\rm exp}\approx
\Delta\sum_{k=1}^{N_{\tau}}g_{k}w_{\tau}(s_{k})v_{1}(s_{k})
\end{equation}
where $N_{\tau}=1+[(m_{\tau}^{2}-s_{1})/ \Delta]_{\rm round}$,
$s_{k}=s_{1}+(k-1)\Delta$ with $s_{1}=0.0125$ and $\Delta=0.025$,
and $g_{k}$ are the numeric coefficients associated with the
trapezoidal rule. From formula (\ref{trpz}) one easily evaluates
the standard error on $J_{\tau,V}^{\rm exp}$
\begin{equation}
\label{sAtau} \sigma(J_{\tau,V}^{\rm
exp})=\Delta\left[\sum_{k=1}^{N_{\tau}}
\sum_{n=1}^{N_{\tau}}g_{k}g_{n}w_{\tau}(s_{k})w_{\tau}(s_{n}){\mathbb
C}_{k,n}\right]^{1\over 2}
\end{equation}
where $\mathbb C$ denotes the covariance matrix ${\mathbb
C}_{i,k}={\bf E}[(v_{1}(s_{i})-\overline
{v_{1}(s_{i})})(v_{1}(s_{k})-\overline {v_{1}( s_{k})}]$ which is
available in \cite{compilation}.  It follows from
Eqs.~(\ref{dQCD}) and (\ref{rate2})  that
\begin{equation}
\label{sdelta0} \sigma(\delta_{\rm
QCD})=[{\sigma}^{2}(\delta^{0})+ \sigma^{2}(\delta_{\rm
NP})]^{1/2}=4\sigma(J_{\tau,V}^{\rm exp}),
\end{equation}
where we have ignored the small correlation between $\delta^{(0)}$
and $\delta_{\rm NP}$. With the data provided by ALEPH
\cite{compilation}, from Eqs.~ (\ref{rate2}), (\ref{sAtau}) and
(\ref{sdelta0}) we obtain \footnote{Alternatively, we could have
determined the error on   $\delta^{(0)}_{\rm exp}$ directly from
the known error on $R_{\tau,V}$ using formula (\ref{rate1}) with
the estimate $\delta_{\rm NP}=0.0199\pm 0.0027$. }
\begin{equation}
\label{exptd0} \label{exp} \delta^{(0)}_{\rm exp}=0.2091\pm
0.0065_{\rm exp},
\end{equation}
it should be noted that  in \cite{DHZ} slightly large value and
error have been obtained,  namely, $\delta^{(0)}_{\rm
exp}=0.2093\pm 0.008_{\rm exp}$. The perturbative QCD correction
obtained within CIPT is represented via the contour integral in
the complex momentum squared  plane \cite{pivo,DP}. This integral
can be rewritten as
\begin{equation}
\label{delta0CI} \delta^{(0)}_{\rm CI}={1\over
\pi}\int_{0}^{\pi}{\rm Re}\left\{\left(1-e^{\imath\varphi}\right)
\left(1+e^{\imath\varphi}\right)^{3} d_{\rm
RGI}(s_{0}e^{\imath\varphi})\right\}d\,\varphi,
\end{equation}
where $s_0=m_{\tau}^{2}$ and $d_{\rm RGI}(z)$ denote the RG
improved perturbative correction to the Adler function defined in
(\ref{RGI_PT}). To calculate $d_{\rm RGI}(z)$, usually,  the
four-loop order RG equation  is solved numerically for the running
coupling. We find  convenient to use  the implicit solution to the
RG equation at the four-loop order (relevant formulas can be found
in \cite{my3}).  The running coupling satisfies a transcendental
equation which is solved numerically. To extract the value of the
QCD scale parameter $\Lambda\equiv\Lambda_{\overline {\rm MS}}$,
one solves the equation
\begin{equation}
\label{extracteq} \delta^{(0)}_{\rm CI}(\Lambda)=\delta^{(0)}_{\rm
exp}.
\end{equation}
Numerical values  for the QCD scale parameter  and   strong
coupling constant (for $n_{f}=3$ number of flavours) extracted
from the experimental value (\ref{exptd0}) are given in Table
\ref{tab:11}. We have used various approximations to the Adler
function evaluated with the four-loop running coupling. For the
unknown $\rm{N}^{4}\rm{LO}$ coefficient of the Adler function, we
have used the geometric series estimate $d_{5}\approx 378\pm 378$
\cite{DHZ}.
\begin{table}
\caption{Numerical values for the QCD scale parameter and strong
coupling constant in the $\overline{\rm MS}$ scheme for three
active flavours extracted from the non-strange vector  $\tau$
lepton  data   within the conventional CIPT approach. The results
obtained in consecutive orders of perturbation theory are given.
The error bars refer to the experimental uncertainty only.}
\label{tab:11}
\begin{tabular}{p{3cm}cc}
\hline\noalign{\smallskip}
Perturbative orders& $\Lambda\,\,\rm{GeV}$&$\alpha_{s}(m_{\tau}^{2})$\\
\noalign{\smallskip}\hline\noalign{\smallskip}
\rm{LO}&$0.604\pm 0.023$& $0.485\pm 0.019 $\\

\rm{NLO}&$0.469\pm 0.018$&$0.390\pm
0.011$ \\

$\rm{N}^{2}\rm{LO}$&$0.430\pm 0.016$&$0.367\pm 0.009$ \\

$\rm{N}^{3}\rm{LO}$&$0.407\pm 0.015$& $0.354\pm 0.008$\\

$\rm{N}^{4}\rm{LO}$&$0.395\pm 0.015$& $0.347\pm 0.008$\\\hline

\noalign{\smallskip}
\end{tabular}
\end{table}

\end{document}